\documentclass[onecolumn,authoryear]{els-mrw} 

\usepackage{amsmath,amssymb,amsfonts,amsthm,makeidx,graphicx}
\usepackage{txfonts}
\usepackage{helvet}

\usepackage{hyperref} 
\hypersetup{hidelinks, colorlinks=true, breaklinks=true, urlcolor= blue, filecolor=blue,linkcolor=blue,citecolor=blue, bookmarksopen=false, pdftitle={Title}, pdfauthor={Author}}

\def\BoxTypeAfont{\fontfamily{\sfdefault}\fontsize{7}{9}\selectfont\parskip=0pt}

\usepackage[dvipsnames]{xcolor}

\usepackage{soul}
\usepackage[utf8]{inputenc}
\usepackage[normalem]{ulem}

\begin{document}

\chapter{Planetary Nebulae}\label{chap1}

\author[1,2]{Orsola De Marco}%
\author[3]{Isabel Aleman}%
\author[4]{Stavros Akras}%

\address[1]{\orgname{Macquarie University}, \orgdiv{School of Mathematical and Physical Sciences}, \orgaddress{Balaclava Road, North Ryde, Sydney, NSW 2109}}

\address[2]{\orgname{Macquarie University}, \orgdiv{Astrophysics and Space Technologies Research Centre}, \orgaddress{Balaclava Road, North Ryde, Sydney, NSW 2109}}

\address[3]{\orgname{Laborat\'{o}rio Nacional de Astrof\'{i}sica},
\orgaddress{Rua dos Estados Unidos, 154, Bairro das Na\c{c}\~{o}es, Itajub\'{a}, MG, CEP 37504-365, Brazil}}

\address[4]{\orgname{Institute for Astronomy, Astrophysics, Space Applications and Remote Sensing}, \orgdiv{National Observatory of Athens}, \orgaddress{15236, Penteli, Greece}} 


\maketitle

\begin{glossary}[Nomenclature]
\begin{tabular}{@{}lp{34pc}@{}}

AGB & asymptotic giant branch\\
ALMA & Atacama Large Millimeter Array\\
au & astronomical unit $=$~1.5$\times$~10$^{11}$~m\\
Chandra & Chandra Space Telescope\\
ELT & Extremely Large Telescope \\
ESA & European Space Agency\\
EXOSAT & European X-ray Observatory Satellite\\
FUSE & Far Ultraviolet Spectroscopic Explorer\\
Gaia & The Gaia space telescope \\
GALEX & Galaxy Evolution Explorer\\
Hipparcos & Hipparcos space telescope \\
HST & Hubble Space Telescope\\
IR & infrared\\
ISO & Infrared Space Observatory\\
IUE & International Ultraviolet Explorer\\
JWST & James Webb Space Telescope\\
L$_\odot$ & Solar luminosity = 3.83 $\times$ 10$^{26}$~W\\
M$_\odot$ & Solar mass = 1.99 $\times$ 10$^{30}$~kg\\
MUSE & Multi Unit Spectroscopic Explorer\\
NASA & National Aeronautics and Space Administration\\
PN & planetary nebula (PNe for the plural form: planetary nebulae)\\ 
PNLF & PN luminosity function\\
pc & parsec = 1.36 light years = 3.09 $\times$ 10$^{16}$~m\\
RGB & red giant branch\\
ROSAT & Roentgen Satellite\\
SKA & Square Kilometer Array\\
SNR & supernova remnant\\
UV & ultraviolet\\
$[$WR] & Wolf-Rayet type of planetary nebula central star\\
Z & the mass fraction of all elements heavier than helium\\

\end{tabular}
\end{glossary}

\begin{abstract}[Abstract]

Planetary nebulae are formed by the matter ejected by low-to-intermediate mass stars ($\sim 0.8-8$ times the mass of the Sun) towards the end of their lives. As hydrogen and then helium fuel sources run out, stars expand. During these giant phases stars also lose sizable amounts of mass. During the second giant phase, after the exhaustion of core helium, the mass loss is so great that stars lose a large fraction of their mass ($50 - 90$\%), leaving behind a small, hot core, known as a white dwarf, surrounded by a nebula. Planetary nebulae are the result of many processes that shape and alter their ionization structure and chemical composition. The resulting nebula, illuminated by the ultraviolet-rich spectrum of the remnant very hot stellar core, is a spectacle of beauty and science. In this chapter, we show that these objects are invaluable laboratories for astrophysics, astrochemistry, and astromineralogy studies, with impact in many areas of Astronomy.

\end{abstract}

\begin{keywords}
Planetary nebulae -- Preplanetary nebulae -- Circumstellar matter -- Stellar evolution -- White dwarf stars -- Stellar winds -- Interstellar gas 
\end{keywords}

\section{Introduction}

Planetary nebulae (PNe) are among the most fascinating objects in the universe. They are gas ejected in the late stages of the life cycle of low-to-intermediate mass stars, such as our Sun. PNe are characterized by stunning, unique shapes and structures, surrounding the remnants of their progenitor star, which after ejecting the nebular gas is now on its way to becoming a compact and hot white dwarf. They provide valuable insight into stellar evolution and the chemical enrichment of the cosmos. In this chapter, we will explore the complexity of PNe and the methods astronomers employ to study their composition and dynamics. Our goal is to present a comprehensive overview of PNe across the electromagnetic spectrum, discuss the physical mechanisms that govern their diverse morphologies, examine the physico-chemical properties of their plasma and molecular gas components, and investigate the evolution of their central stars, whether they are single or binary systems.

\section{A Modern Introduction to Planetary Nebulae} \label{sec:the_evolutionary_path_to_planetary_nebulae}

\begin{figure}[h!]
    \centering
    \includegraphics[width=9.0cm]{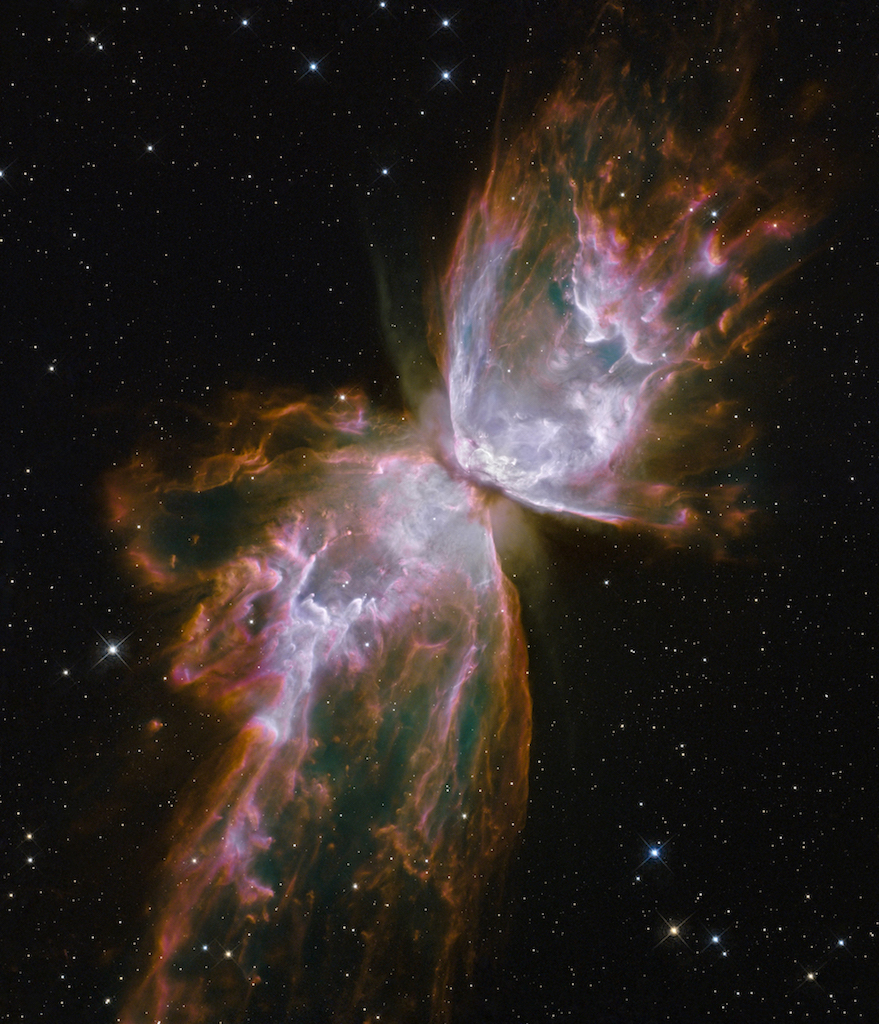}
    \caption{The planetary nebula NGC~6302 observed by the Hubble Space Telescope Wide Field Camera. This nebula is approximately 3300 light years away and is approximately 3 light years across. Colors are white for [S~{\sc ii}], orange for [N~{\sc ii}], brown for H$\alpha$, cyan for [O~{\sc iii}], blue for He~{\sc ii}, and purple for [O~{\sc ii}]. Credit: NASA, ESA.} 
    \label{fig:PN}
\end{figure}

The classic picture of stellar evolution that leads to the production of planetary nebulae (PNe) has been in place for many decades and its general guidelines, explained below, continue to hold. That said it has recently become clear that the events that shape these spectacular astronomical objects (see Figure~\ref{fig:PN}) are far more complex and connected to many classes of different astronomical objects and phenomena, from supernovae, to planet formation and demise, to the makeup of galactic populations. In this chapter we will explain how PNe keep Astrophysicists busy and interested and the general public entertained and amazed.

\begin{figure}[h!]
    \centering
    \includegraphics[width=12.0cm]{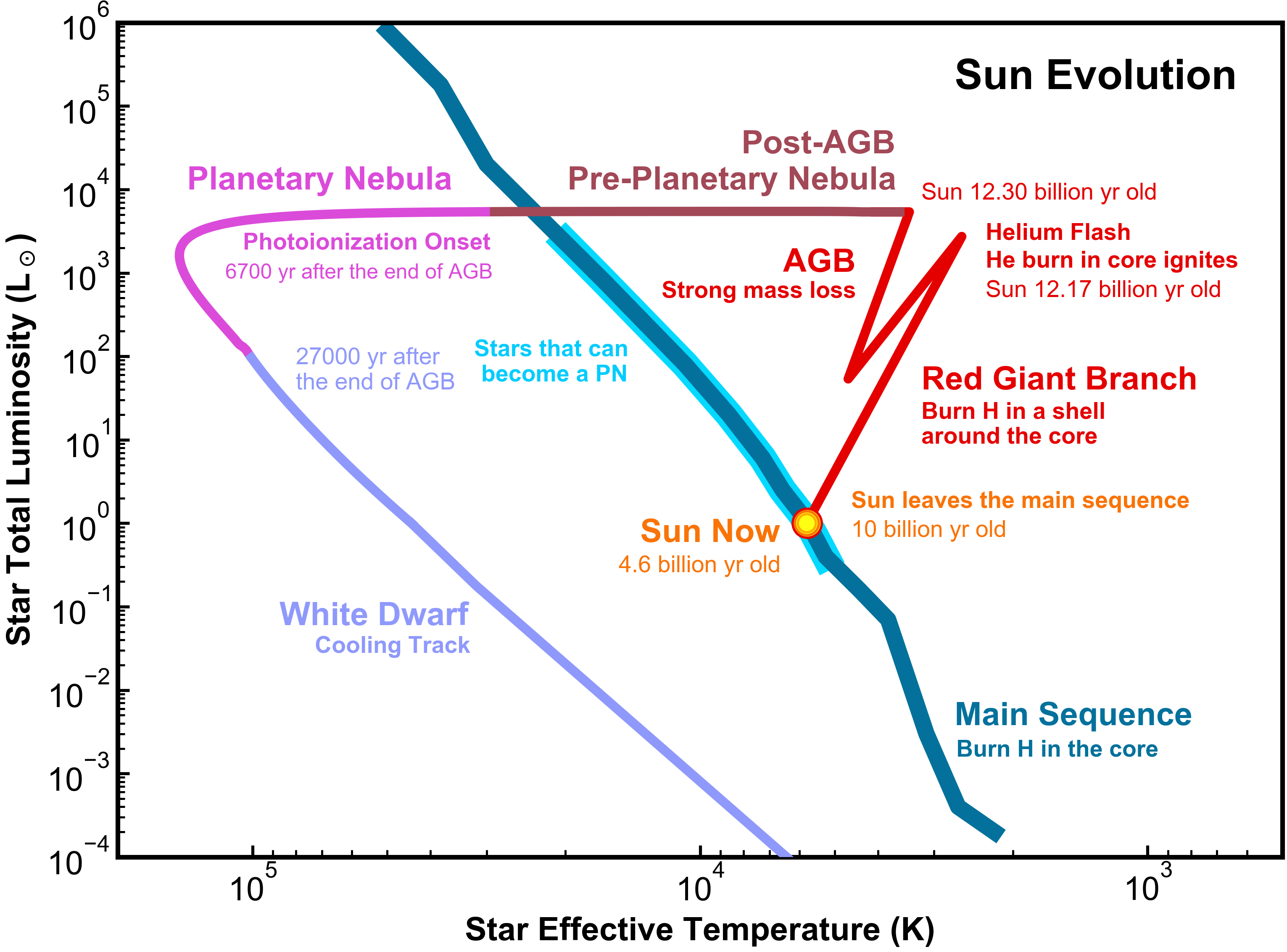}
    \caption{The Hertzsprung-Russell diagram for the Sun, a representation of the change in effective temperature and luminosity over the Sun's life, from today (``Sun Now") to when it will become a white dwarf. The thick blue line is instead the \lq\lq main sequence\rq\rq, the locus where stars that burn hydrogen in their core, with lower mass stars occupying the bottom right side and more massive stars the top left. The red line is the Sun's evolution  till the end of the AGB phase, the brown line is the post-AGB, pre-PN phase, the pink line is the PN phase while the purple line is the time when the PN has faded and the star cools as a white dwarf.}
    \label{PN_evol_new}
\end{figure}

PNe are formed at the end of the life of low-to-intermediate mass stars ($\sim 0.8-8$~M$_\odot$; Figure~\ref{PN_evol_new}). Such stars spend several billion years converting hydrogen into helium. The very energetic nuclear fusion reaction heats the stellar gas and the resulting pressure balances the gravity resulting in a long phase of equilibrium known as the stellar \lq\lq main sequence\rq\rq\ (blue line in Figure~\ref{PN_evol_new}). Our Sun is currently halfway through its 10 billion years on the main sequence. Once the hydrogen in the center of these stars starts to run out, the balance is altered and the star needs to adjust its characteristics, such as its size, to a new equilibrium. As the star expands and cools, we say that it \lq\lq leaves the main sequence\rq\rq, and becomes a giant. The first time giant is known as a red giant branch (RGB) star. At some point, helium, now plentiful in the stellar core, ignites (either in a flash or more quietly) giving the star a new lease on life, as a core-helium burning star (yellow symbol labeled as ``yellow giant" in Figure~\ref{PN_evol_new}). Once helium runs out, having fused to carbon and oxygen, a new expansion sees the star grow for the second time, becoming, this time, an asymptotic giant branch (AGB) star. This phase leads to very cool outer layers where molecules and dust can form. Through an as yet poorly understood mechanism, radiation on dust promoted by stellar pulsations, pushes mass out, till the large stellar envelope becomes too light to retain its size. The star shrinks in size with a resulting heating of the photosphere. This hot central star emits abundant high energy photons. These photons first dissociate the molecular gas in the now ejected and expanding circumstellar gas and then ionize and heat the atomic gas. The PN phase begins with the onset of the photoionization of the gas and lasts on the order of tens of thousands of years till the nebular gas disperses and the fading star fails to keep its gas ionized (for a review see \citealt{Iben1995}).

The elemental and chemical composition of PNe can be easily measured even for trace elements thanks to their brightness. Some elements are the result of the elemental creation in the stellar furnace, and give us clues on how stars that make PNe work. Other elements are generated by different processes, e.g., they are created in much more massive stars, and ejected later into space. They are then  inherited by the star that becomes the PN central star, from the interstellar gas it forms from. When we measure them in the nebula we understand how the makeup of interstellar gas changes over time due to the combined action of several processes. As the central star is heating up over time, the nebular gas is subject to change: it can be ionized, it can recombine, it can associate into molecules or build solid grains. As such PNe serve as unique astronomical laboratories, providing insight into plasma (i.e., ionized gas) physics, molecular processes, dust formation, as well as stellar nucleosynthesis, chemical enrichment of the interstellar medium and the evolution of galaxies.

While the star is on the AGB it ejects its envelope in a dense, slow wind. Following the ejection of a significant portion of its mass, the exposed core starts to blow a much thinner but faster wind. The interaction between these two winds plays a crucial role in shaping the structural characteristics of PNe \citep{Kwok_1978}. 

For the longest time the cause of bipolar symmetry and other types of asymmetry was ascribed to stellar rotation and magnetic fields \citep[e.g.,][]{GarciaSegura1999,Balick_Franck_2002_review}. However, it was eventually realized that such physical mechanisms are not likely to happen in single stars \citep{Nordhaus2006,Soker2006,GarciaSegura2014}, thus binary interactions between the AGB star and one or more companion stars, or even planets, became a preferred channel for the formation of non-spherical PNe \citep{DeMarco2009,Jones2017b}. This realization also meaningfully connected PNe to the diverse group of transients, astronomical objects whose brightness varies on relatively short timescales \citep{Kasliwal2011}. Many of these are outbursts and explosions due to binary star interactions. In this scenario, the binary interaction that ejected and shaped the PN would have been observed as a transient event in the past. 

Because of their characteristic timescales (short, ensuring that the ejection was recent, yet not too short, resulting in sufficient numbers for reasonable statistical studies) PNe are phenomenal laboratories to study stellar evolution, binary interactions, the origin of the elements (nucleosynthesis), formation and evolution of dust, stellar populations, galactic chemical and dynamical evolution and more. In what follows we will describe some of these studies and the results that have been obtained over the last few decades.

\section{Morphologies, Structures and Classifications}\label{sec:nebularstructres}

\begin{figure}[h!] 
    \centering
    \includegraphics[width=\textwidth,trim={0.0cm 0.0cm 0.0cm 0.0cm},clip]{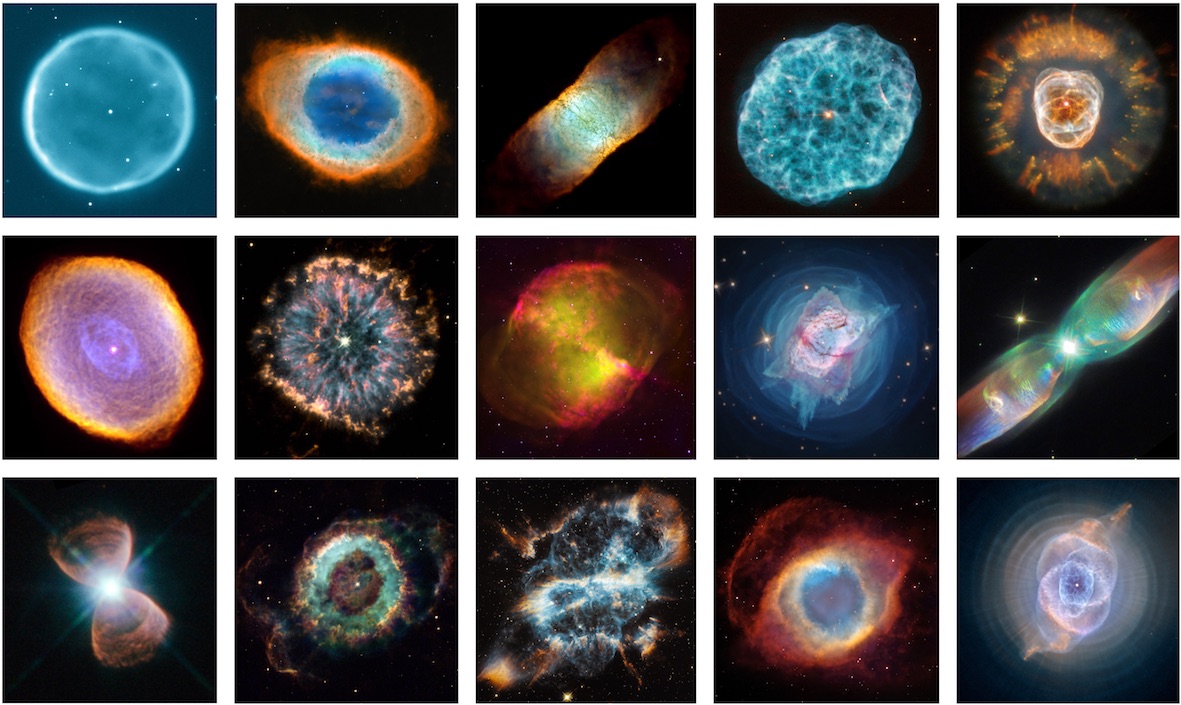}
    \caption{Observed planetary nebulae with different shapes. Credits for the PN images: NASA, ESA, HEIC, WIYN, NOIRLab, NSF, The Hubble Heritage Team STScI/AURA, the ERO team (STScI + ST-ECF), C. R. O’Dell, J. Kastner, K. Noll, J. Schmidt, M. Canale, A. Fruchter, M. Meixner, P. McCullough, G. Bacon, G. Jacoby, J. Barrington, and A. Zijlstra.}
    \label{fig:images_PNe}
\end{figure}

From a morphological point of view, PNe are extremely diverse. They exhibit a variety of shapes and structures that produce beautiful visual displays (Figure~\ref{fig:images_PNe}). Based on the observed  shapes, they can be categorized into five distinct subgroups: round, elliptical, bi (or multi) polar, point symmetric and asymmetric \citep{Balick_1987,Balicketal1987,Manchado_etal_1996,Sahai_2007_classification}. It is important to realize that what we see is a projection of the three-dimensional PN  structure on the plane of the sky and inferring their real, 3D structure is not straightforward. For example, elliptical or bipolar PNe may be erroneously classified as spherical, as they can present as circular on the sky if viewed down their symmetry axis, and multipolar PNe can result in more intricate projected shapes \citep{Chong_tal_2012}.

\begin{figure}[h!]
    \centering
    \includegraphics[height=5.5cm]{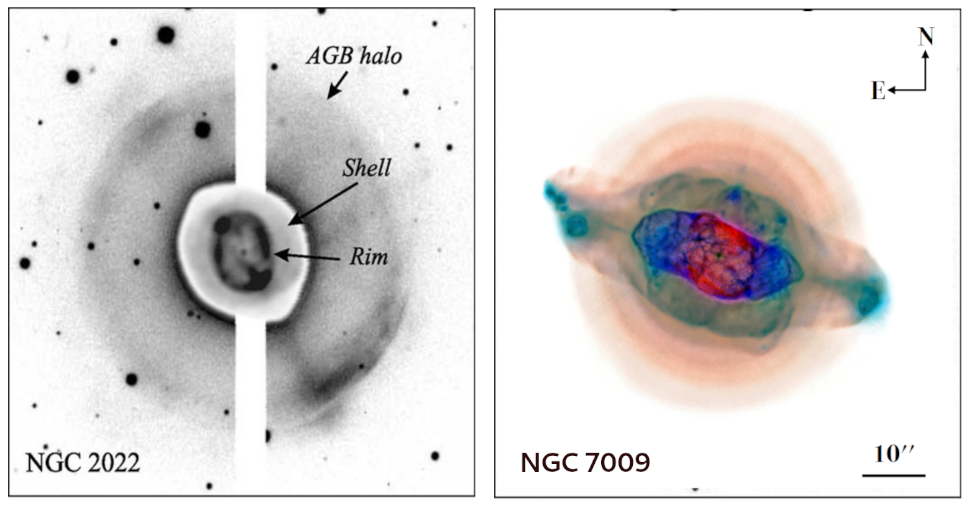}
    \caption{Main structures often seen in a planetary nebula. Left panel: the rim, shell and the much fainter halo nebular structures are indicated on the optical [O~{\sc iii}] emission line image of NGC~2022 \citep[adapted from figure~1 of][]{Corradi_etal_2003}. The central part of the image (shell and rim) and the outer part (halo) are shown in different intensity scales (separated by the white gap) to better represent the very different emission intensities. Right panel: concentric rings are observed in the halo of NGC~7009 \citep[adapted from figure 1 of][]{Guerrero_etal_2020}.}
    \label{rimshellhalorings_PNe}
\end{figure}

We typically distinguish three, large scale PN structures: the rim, the shell and the halo (Figure~\ref{rimshellhalorings_PNe}). They are usually identified from images taken using filters that isolate the light of ionized hydrogen (e.g., H$\alpha$ and H$\beta$ in the optical wavelength range) as well as that of bright emission from twice ionized oxygen (O$^{++}$) and singly ionized sulfur (S$^{+}$). The rim is a narrow region of shocked AGB star gas, swept up by the faster post-AGB wind. The shell is the thermal pressure-driven ionized AGB material. The halo is the much fainter ($\sim$1000 times) AGB material, most commonly spherical, that surrounds the shell and rim structures. It is typically a few times larger than the shell \citep[e.g.,][]{Balick_etal_1992, Corradi_etal_2003}. 

In 2000, high-resolution Hubble Space Telescope (HST) images of PNe revealed the presence of faint concentric rings in the halos of four PNe \citep{Terzian2000}. The right panel in Figure~\ref{rimshellhalorings_PNe} displays the rings (pink structures) around the well-known Saturn nebula \citep{Guerrero_etal_2020}. Later, \cite{Corradi2004} carried out a systematic search for rings in PNe and managed to triple the number, concluding that the separation between the rings varies from 0.005 to 0.06~pc and the processes producing the rings should be repeated every 500 to 4000 years \citep{Corradi2004}. \cite{Phillips2009} reported the first detection of rings in mid-IR observations from the Spitzer Space Telescope. A search for rings in the HST and Spitzer image archives led to 29 new detection \citep{RamosLarios2016}. These high quality data allowed them to get better estimates of the time between successive rings, namely between 500 and 1200~years \citep{RamosLarios2016,Guerrero_etal_2020}. Last but not least, the high quality ``integral field unit" data of the PN IC~4406 allowed us to perform the first spectroscopic analysis of the rings \citep{RamosLarios2022}, that turned out to be characterized by a lower ionization gas, a comparable electron temperature, T$_{\rm e}$, but lower electron density, n$_{\rm e}$, than the rim.

These rings are likely imprints of orbital motion, where a companion either carved a spiral in the outflowing AGB wind, or induced periodic reflex motion of the AGB star itself. Spiral structures tend to become ring structures at some distance from the source \citep{Decin2021}. Typical ring separations and adopted wind velocities can be translated into orbital separations of 50-200~au.

\begin{figure}[h!]
    \centering
    \includegraphics[height=9.0cm]{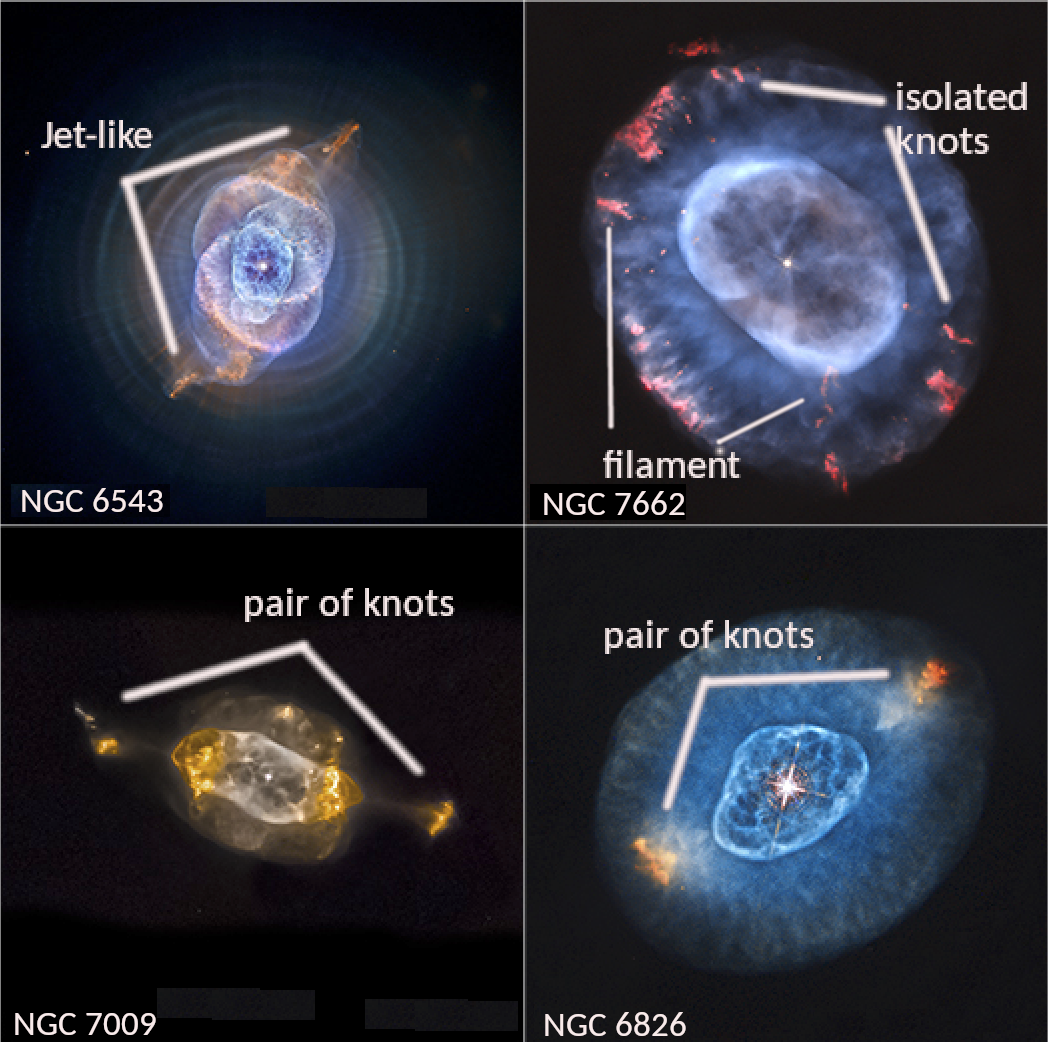}
    \caption{Four PNe with different types of small scale structures obtained from the Hubble Space Telescope archive.}
    \label{LISstypes_PNe}
\end{figure}

PNe also exhibit small scale structures in the form of knots, filaments, and jet-like features (Figure~\ref{LISstypes_PNe}). These small scale structures were first discovered in high resolution, narrow-band images in the light of low ionization emission species of oxygen (at $6300~\AA$) and or singly ionized nitrogen (at $6584~\AA$; \citealt{Balick_1987,Balick1993,Corradi1996}). For this reason, they are also known as low-ionization structures. Low ionisation structures have slightly lower, but overall similar electron densities to the surrounding nebular gas, while no differences in electron temperature or chemical composition have been found \citep[e.g.,][]{Goncalves2001,Mari2023}. Some PNe, mostly elliptical objects, exhibit pairs of low ionisation structures positioned on opposite sides along the main symmetry axis (Figure~\ref{LISstypes_PNe}). Such structures have velocities $\sim$50~km~s$^{-1}$, much higher than the typical PN expansion velocities ($\sim$20~km~s$^{-1}$). These sub-structures are sometimes known as fast, low-ionization emission regions \citep[][]{Balick1993,Balick1994}.

\begin{figure}[h!]
    \centering
    \includegraphics[height=10.0cm]{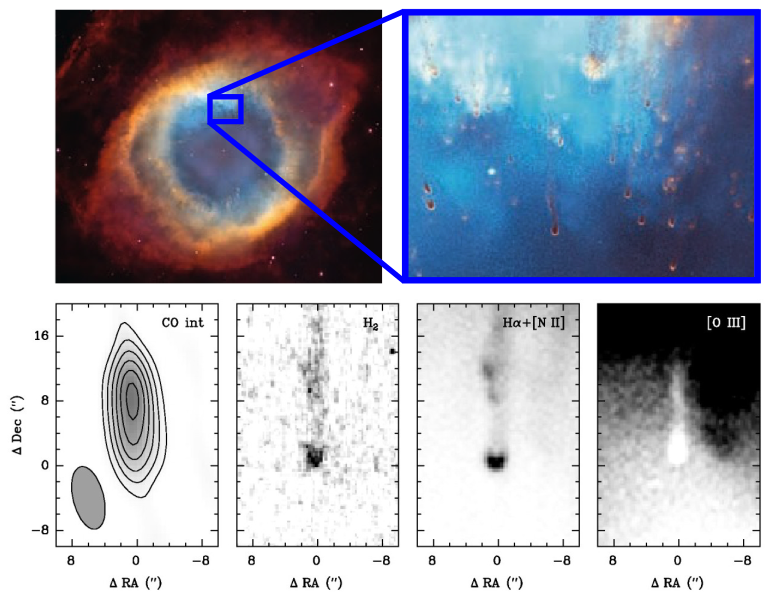}
    \caption{Top left panel: the cometary knots in the Helix PN. Top right panel: enlargement of a detail from the left panel shows that the cometary knots form part of the helicoidal structure that gives this PN its name. Figure adapted from NASA, NOAO, ESA, M. Meixner and T. A. Rector and \citet{Matsuura_etal_2009}. Bottom panels: one cometary knot is imaged at different wavelengths, showing ionic and molecular emissions. Figure adapted from figure 1 of \citet{Huggins_etal_2002}.}
    \label{fig:helix_knots}
\end{figure}

Molecular emission (H$_2$, CO) was detected with the Atacama Large Millimeter Array (ALMA) in the cometary knots, initially of the Helix and Ring PNe \citep{Huggins_etal_2002, Speck_etal_2003, Matsuura_etal_2009, Andriantsaralaza_etal_2020}. These cometary knots are PN microstructures with a round head and a tail (Figure~\ref{fig:helix_knots}), a shape reminiscent of comets, thus their name. Recently, line emission from H$_2$ has also been detected in other kinds of low ionisation structures, further indicating that these structures are dense clumps of low ionization gas, within which molecular gas can survive because it is shielded from the radiation of the hot star by the dense material \citep{Akras_etal_2017, Akras_etal_2020}.

\section{Observing and Modeling the Nebulae}
\label{sec:observing_and_modeling_planetary_nebula}

To first approximation, PNe are simple objects: a central radiation source ionizes the surrounding gas, which is composed mainly of hydrogen (which constitutes one in ten of all particles). The remaining gas is composed of helium, with only a trace of other elements (often referred to as ``metals'' in astronomy). Central stars of PNe have temperatures in the range 30\,000--300\,000~K and emit numerous high energy photons, particularly in the ultraviolet, which interact with the surrounding matter. The photons are absorbed and scattered by the gas and dust particles, ionizing atoms, exciting energy levels and heating dust grains. 

Close to the star, ultraviolet photons keep the nebular gas fully ionized.  For each element, an onion-like structure is formed, with the more ionized ions close to the star and the less ionized ions in the outer zones. The number of photons emitted by the star can only ionize a limited quantity of gas. If the nebula has more gas than that, the ionized region will be surrounded by a neutral region and farther out, an outer molecular shell. Since the gas is not uniformly distributed (Figure~\ref{fig:images_PNe}), these concentric regions can diverge from shells, but the general idea remains.

In addition to UV heating via the stellar radiation, there is a secondary, lesser, contribution from shocks. Nebular gas moves slightly supersonically (with typical velocities of $20-30$~km~s$^{-1}$ that are just faster than the local sound speed, which, in a gas with $T_\textrm{e} \sim$~10\,000~K is $\sim 10-15$~km~s$^{-1}$). However, faster moving gas with velocities $> 40-50$~km~s$^{-1}$ results in shocks that can considerably contribute to the nebular heating. This shock interaction can be noticed by the increase in $T_\textrm{e}$, which also leads to the enhancement of the [O~{\sc iii}] emission line. Several structural components in PNe have been found to exhibit high ratio of the [O~{\sc iii}] line at 5007~\AA\ and the H$\beta$ line, revealing the contribution of shock heating \citep{Guerrero2013}. We further note that the bright rim often observed in PNe (e.g., in Figure~\ref{rimshellhalorings_PNe}) is also the result of shock interaction between the fast stellar wind from the post-AGB phase, which swept out the material from the AGB phase. A reverse shock is also formed. It moves inwards and interacts with the stellar wind, resulting in the formation of a hot bubble that shines in X-rays (see Section~\ref{sec:uv_and_xrays_emission}).

Dust particles are effective in absorbing and scattering UV and optical photons. Absorbed photons can heat the dust particles, which, in turn, re-emit part of this energy back in the IR range. UV photons also cause photoelectric effect on the grain's surface. This can be an important gas heating mechanism in the lower ionization and neutral regions.

\begin{BoxTypeA}[nebular_physics]{\normalsize \usefont{OT1}{phv}{m}{n}\selectfont{Nebular Physics}} \label{box_neb_phys}

\section*{\normalsize \usefont{OT1}{phv}{m}{n}\selectfont{Photoionization}}

Photoionization occurs when an electron in an atom or molecule absorbs a photon with enough energy to eject it. The energy must be larger than the ionization potential for that electron's energy level. UV light can usually eject electrons from the outer electronic shells of atoms and molecules and it is therefore considered ionizing radiation.

\section*{\normalsize \usefont{OT1}{phv}{m}{n}\selectfont{Recombination}}

Recombination is the (free-bound) process whereby an electron is captured by an ion and a photon (or sometimes two) is emitted. Electrons typically recombine into excited energy levels, but then lose energy and ``cascade" down to lower energy levels (bound-bound process), emitting additional line photons.

\section*{\normalsize \usefont{OT1}{phv}{m}{n}\selectfont{Bremsstrahlung}}

In ionized regions, the gas is dominated by positively charged protons and negatively charged electrons deriving from hydrogen atoms (there may be also ions of heavier elements where one or more electrons have been stripped off atoms). In this gas, electrons move among the protons and are either accelerated or decelerated. In the former case they absorb energy in the form of photons, in the the latter, they emit them. This radiation does not have a specific wavelength (free-free process), so it is said to be ``continuum'' radiation.

\section*{\normalsize \usefont{OT1}{phv}{m}{n}\selectfont{Other Mechanisms}}

In the ionized region of a PN, photoionization and recombination are the most important mechanisms determining the ionization degree of atomic species. Other physical processes and chemical reactions can happen inside the nebula, involving photons, atoms, molecules and dust. Their importance will depend on the physical conditions (temperature, density, or radiation field) and the inherent likelyhood of that process taking place.

\section*{\normalsize \usefont{OT1}{phv}{m}{n}\selectfont{Ionization and Chemical Equilibrium}}

Ionization equilibrium is established when the number of ionizations (per unit of volume and per unit time) is equal to the number of recombinations (per unit of volume and time). Here all the processes that ionize and recombine must be taken into account for a given species

The chemical equilibrium, a more general case, occurs when all the reactions that form a species balance all the the reactions that destroy it (per unit of volume and per unit time). In addition to ionization and recombination processes, other kinds of reactions may occur. 

\section*{\normalsize \usefont{OT1}{phv}{m}{n}\selectfont{Heating Mechanism}}

A heating process that increases the nebular electron temperature, increases the average of the electron energy distribution. The major source of heating in the ionized zone of a PN is the ionization of hydrogen and helium, which releases electrons with energies of a few electron volts.

\section*{\normalsize \usefont{OT1}{phv}{m}{n}\selectfont{Cooling Mechanism}}

A cooling process reduces the nebular electron temperature by removing energy from the electron energy distribution. In PNe, hydrogen and helium electronic recombinations are important cooling mechanisms. Another significant cooling mechanism is the collision between ions and electrons, which converts kinetic energy into potential energy of the ions (electrons transition to a higher energy state within the ion), which is then lost via photons when the electrons transition downwards again.

\section*{\normalsize \usefont{OT1}{phv}{m}{n}\selectfont{Thermal Balance}}

The nebular electronic temperature is determined by the balance of the energy gain (heating) and loss (cooling) by the electrons. If the thermal equilibrium is assumed, the total energy gain (per unit of volume and time) is equal to the total energy loss (per unit of volume and time) by the electron gas. 

\section*{\normalsize \usefont{OT1}{phv}{m}{n}\selectfont{Optically Thick and Thin Nebulae}}

When photons cannot escape the gaseous nebula, the nebula is defined as ``optically thick" (or ionization bounded, meaning that nebular gas may reside outside the visible edge of the nebula). If photons can escape, the nebula is said to be ``optically thin" (or matter bounded, meaning that no nebular gas resides outside the visible boundary of the nebula). As the interaction of photons with matter depends on their energy, a PN may be optically thin for some photons and thick for others. Also, a PN gas distribution may vary with the direction and therefore the PN may be optically thin in one direction and thick in another, for the same photon energy.

More details on the physics of gaseous nebulae can be found in the book by \citet{Osterbrock_Ferland_2006}.

\end{BoxTypeA}

\subsection{Nebular Gas: Diagnostics and Composition}\label{sec:plasmas}

The analysis of a PN spectrum (particularly from UV to IR) shows many emission features produced by ionized gas. The characteristics of the PN emission line spectrum depend on its composition (i.e., the abundance of elements) and physical conditions (e.g., ionic densities and gas temperature). As mentioned above, this gas is the nebular gas composed by the material ejected by the progenitor star during its previous phases, which is ionized and heated by the very hot remnant stellar core (the central star). 

\begin{BoxTypeA}[emission_lines]{\normalsize \usefont{OT1}{phv}{m}{n}\selectfont{Emission Lines in a PN}}

\section*{\normalsize \usefont{OT1}{phv}{m}{n}\selectfont{Recombination Lines}}

The spectrum of a PN, from UV to radio wavelengths, displays many emission lines of hydrogen.
These lines are produced by the downward transition of electrons through the levels of hydrogen, following the recombination of an electron with the hydrogen ion ($H^+$), thus the name `recombination lines'. In the recombination process, the ionized hydrogen atom captures a passing electron. The electron recombines to any bound quantum state (free-bound electronic transition), producing a ``continuum" photon, so called because these photons can have any energy. If the recombination is to an excited state, the electron progressively decays to less excited levels, emitting photons with well determined energies or wavelengths (bound-bound transitions). The line photons produced in these bound-bound transitions are called the recombination lines.

Electron transitions to the lower level with principal quantum number n=1 produce Lyman line photons, which are mostly in the UV. If the level has principal quantum number n=2, they are called Balmer lines and the more prominent ones are in the optical range (Figure~\ref{fig:optical_Tc1}). Additional series of lines are in the IR. The lines in the Balmer series are called H$\alpha$, H$\beta$ and so on, in on order of decreasing wavelengths. Being in the optical range they are often observed and used to measure different properties of the PN gas. Recombination lines of helium and a few other heavier elements can also be observed. The latter are usually much fainter, as the heavy elements are much less abundant than hydrogen and helium.

\section*{\normalsize \usefont{OT1}{phv}{m}{n}\selectfont{Forbidden or Collisionally Excited Lines}}

The so-called ``forbidden'' lines are produced by the radiative decay in low probability transitions, which, under typical laboratory conditions, would happen much faster by collision with another particle (which does not produce a line). The historical name ``forbidden'' (chosen as they violate the parity selection rule for the most probable electric dipole transitions) is not strictly correct, as the spontaneous transition for these lines is not strictly forbidden, just very unlikely to occur under laboratory conditions. At the low densities typical of PNe, the chance of collisions is much lower than in the laboratory and these lower probability transitions  can dominate. Typical PN gas has a density thousands of times lower than the best vacuum produced in laboratories. These lines are also called \lq\lq collisionally excited\rq\rq\ lines, as the radiative excitation of the upper level is also unlikely, thus happening more often via collision with free electrons in the gas. Forbidden lines are usually denoted by brackets around the ion spectroscopic notation. For example, [O~{\sc iii}] at 4363~$\AA$ and [Ne~{\sc ii}] at 12.8~$\mu$m are typically very bright in PNe.

There are also semi-forbidden lines, that violate the spin selection rule for electric dipole transitions, which also have low probability of radiative transition, but not as low as forbidden lines. Semi-forbidden lines are represented with one bracket on the right side of the ion spectroscopic notation, for example the C~{\sc iii}] line at 1909~\AA.
Although the abundance of metals is much lower than that of hydrogen, some metal forbidden lines are as intense or more intense than the brightest hydrogen lines, and metal forbidden lines in PN  tend to be much stronger than metal recombination lines.

\end{BoxTypeA}

\begin{figure}[h!]
    \centering
    \includegraphics[width=7.5cm]{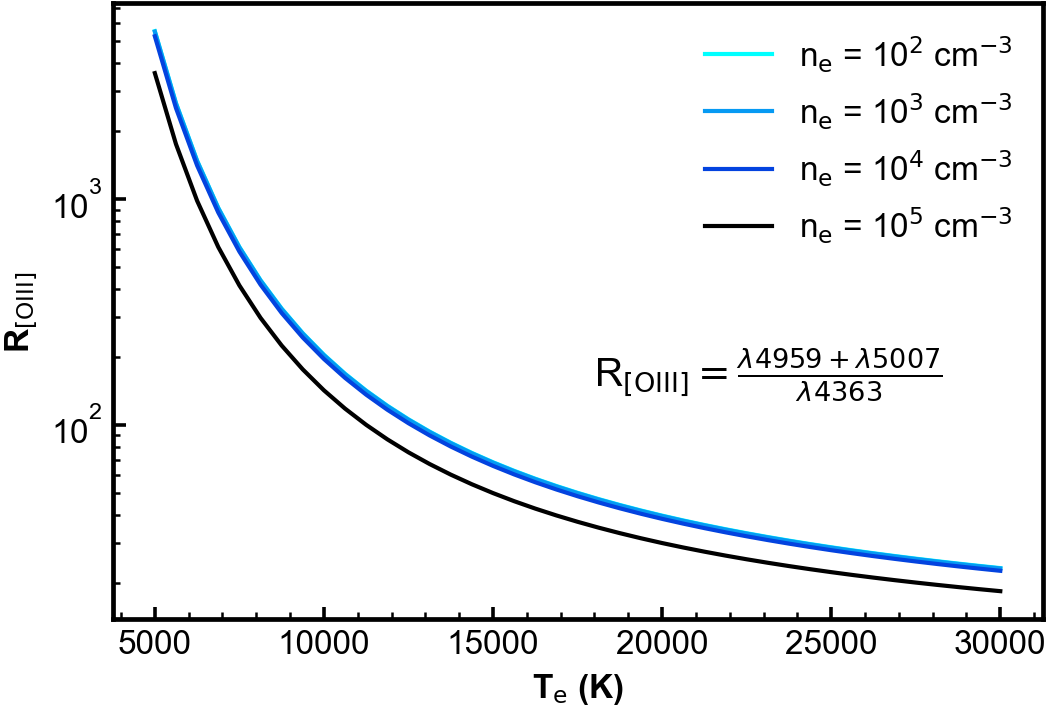}
    \includegraphics[width=7.5cm]{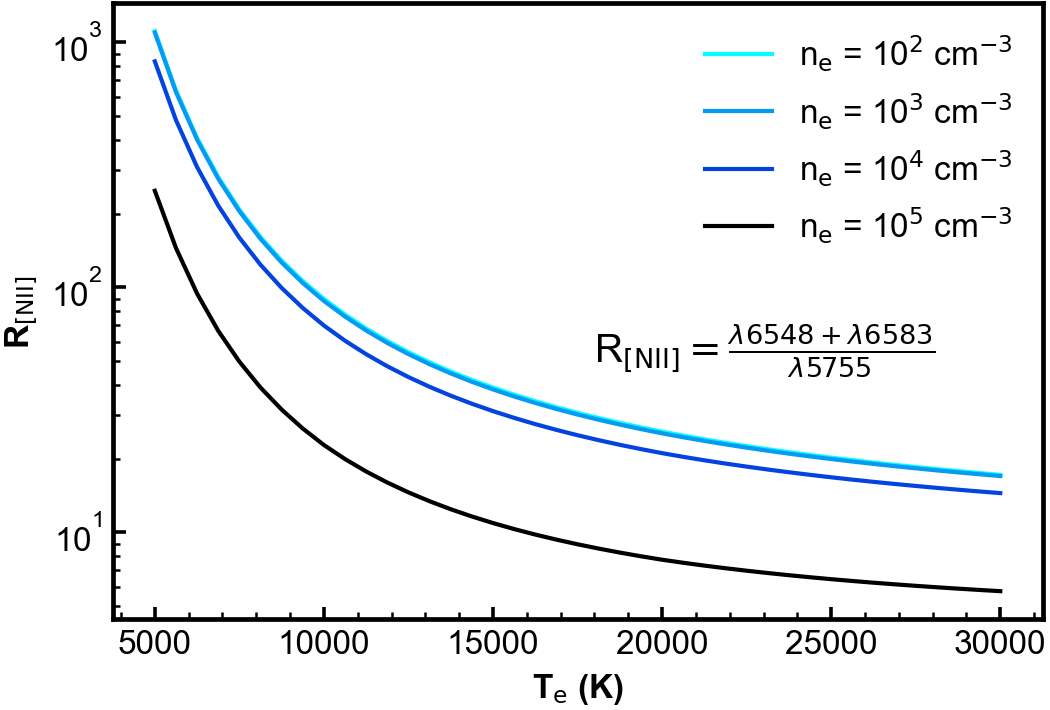}
    
    \includegraphics[width=7.5cm]{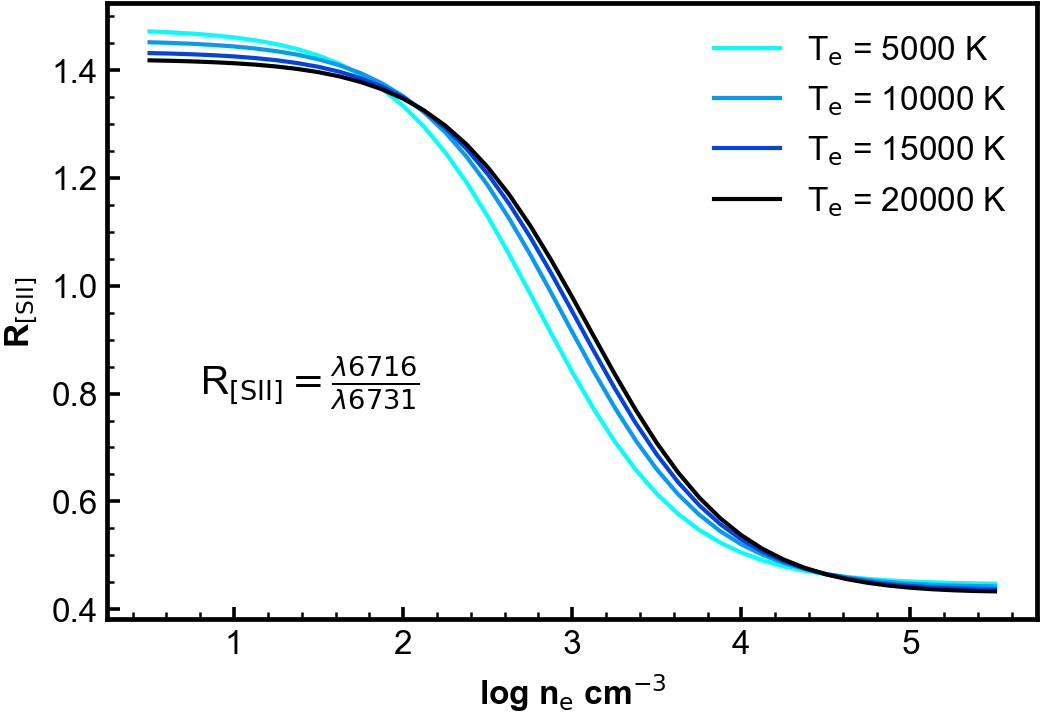}
    \includegraphics[width=7.5cm]{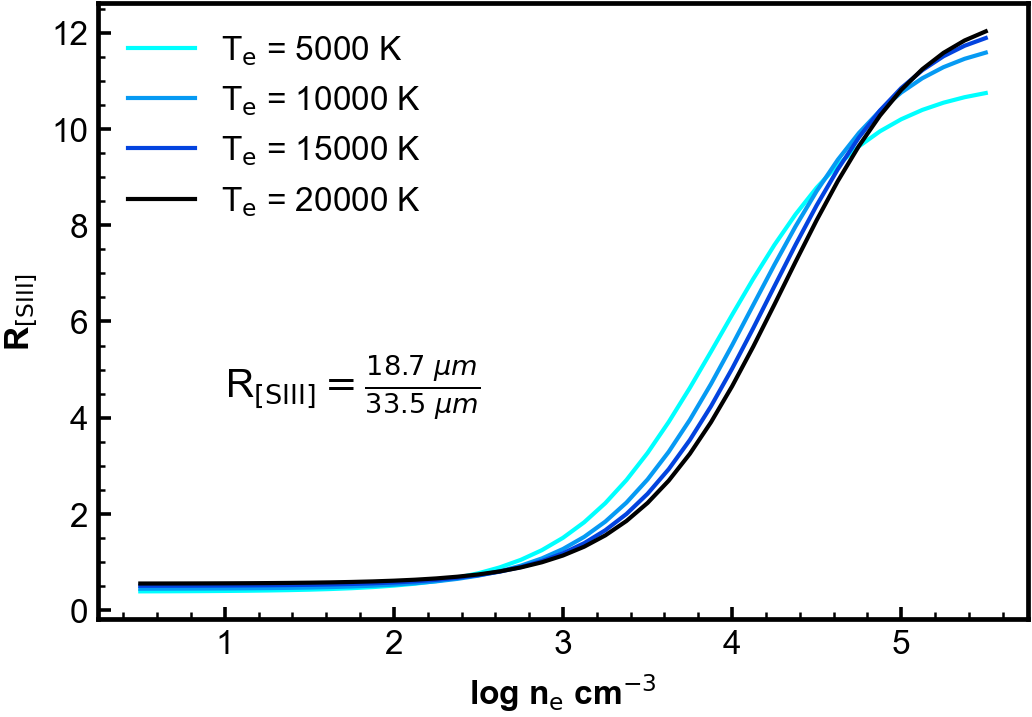} 
    
    \caption{Examples of modeled line ratios as a function of PN gas temperature and density, used as physical condition diagnostics in PNe \citep[
    ][]{Luridiana_etal_2015_PyNeb}.}
    \label{TeNeplots}
\end{figure}

There are several emission lines used for the determination of the electron density and temperature of the gas. In the optical spectral regime, the most common electron temperature diagnostics are the [O~{\sc iii}] and [N~{\sc ii}] lines and the electron density diagnostic are the [S~{\sc ii}] lines (Figure \ref{TeNeplots}). Both temperature diagnostics are valid within a wide range of $T_\textrm{e}$ (5000-20000~K) and $n_\textrm{e}$ (10$^2$-10$^5$ cm$^{-3}$). The ratio of the [S~{\sc ii}] lines at 6716~\AA\ and 6731~\AA\ can measure electron densities in the range  $n_\textrm{e}$=10$^2$-10$^4$~cm$^{-3}$ and $T_\textrm{e}$=5000-20000~K. It should also be noted that the emission lines [N~{\sc ii}] at 5755~\AA\ and [O~{\sc iii}] at 4363~\AA\ may also be affected by recombination emission and this contribution must be corrected for in order to get reliable electronic temperatures \citep[e.g.,][]{Liu2000b,GomezLlanos2020,GarciaRojas2022}. A comprehensive review of nebular diagnostics can be found in \citet{Peimbert_etal_2017_Nebular_Spectroscopy}.

The nebular gas is also characterized by a particular composition, which is associated with the composition of the gas when the progenitor star was formed as well as the chemical elements formed by nuclear fusion reactions in the interior of the star during its evolution. The ionic abundances of different elements such as hydrogen, helium, carbon, oxygen, nitrogen, argon, sulfur, and neon are determined using measurements of collisionally excited and recombination emission lines.

Unfortunately, it is common that no lines of one or more abundant ions are observed. The missing ionic abundance values cause an underestimation of the total elemental abundance (which is the sum of the abundances for all the ions of a given species). The solution is to apply a correction for the unobserved ions, the so called ``ionization correction factor". Widely used sets of correction factor expressions are from \cite{KingsburghandBarlow1994} and \cite{DelgadoInglada2014}. It is important to mention that these correction factors were developed for use with integrated spectra, where the entire PN is sampled and may not be applicable for spectra of small regions of a PN such as those obtained from ``integral field unit" spectroscopic observations. 

Abundances derived from recombination and collisionally excited lines for the same element do not always agree for PNe or H~{\sc ii} regions (regions of ionized hydrogen around newly born stars). Values derived from recombination lines are often higher than those derived from collisionally excited lines. The reason for this difference is still not understood. The ``abundances discrepancy factors", i.e., the ratio between the abundances derived from recombination lines and collisionally excited lines, can be as high as 800 in extreme cases, while values for typical PNe are in the 2-5 range \citep{Peimbert_etal_2017_Nebular_Spectroscopy}. Recent studies have shown that there may be a link between stellar duplicity and the discrepancy factor \citep{Corradi2015, Wesson2018}.

\subsection{Planetary Nebulae Across the Electromagnetic Spectrum} \label{ssec:planetary_nebulae_across_the_spectrum}

\subsubsection{Optical Emission} 
\label{sec:optical}

PNe were first observed because of their optical emission. Charles Messier observed the Dumbbell nebula (M~27 in his famous catalog, \citealt{Messier_1781}) in 1764. The first PN observations showed a fuzzy, roundish, and ``greenish-blue'' \citep{Huggins_Miller_1864} object that reminded the first astronomers of a planet seen through a telescope, hence the misleading name we still use \citep{Herschel_1785}. The first PN spectrum was obtained by \citet{Huggins_Miller_1864} and revealed the now familiar set of emission lines with virtually no continuum emission. This observed spectrum resolved a long standing discussion about the nature of PNe, and their distinction from stars and other ``nebulae'' made of stars (what we now know  to be star clusters or galaxies).

\begin{figure}[h!]
    \centering
    \includegraphics[width=\textwidth]{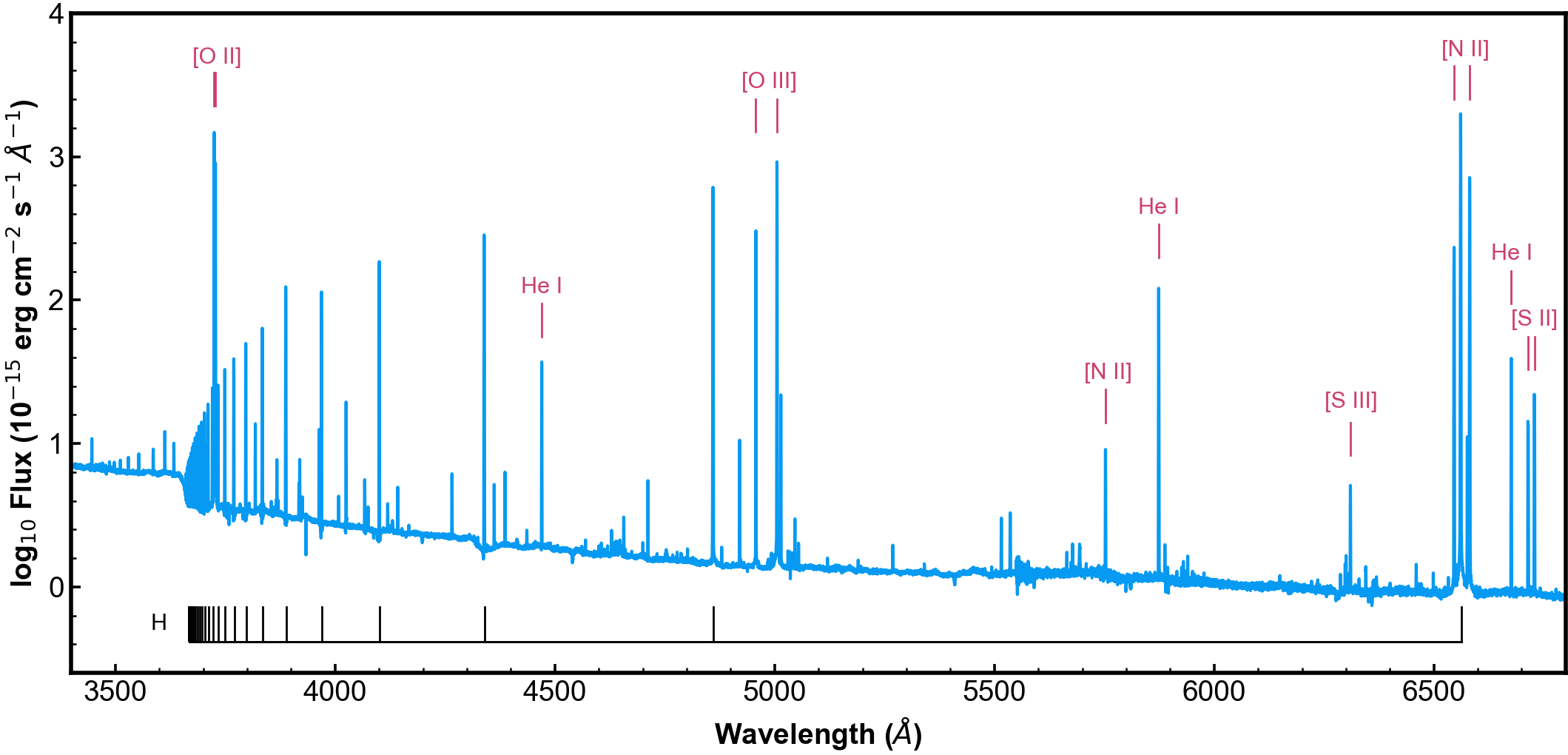}
    \caption{Part of the optical spectrum of the PN Tc~1 obtained with the Very Large Telescope X-Shooter spectrograph \citep{Aleman_etal_2019_Tc1}. The brightest lines are identified. The wavelengths of the hydrogen Balmer series lines are indicated in black. H$\alpha$ (the strongest of the Balmer series) is the rightmost line, with H$\beta$, H$\gamma$, etc. to its left, in alphabetical order. The Balmer ``jump'' can be seen as an abrupt downward jump in the continuum emission at 3647~$\AA$.} 
    \label{fig:optical_Tc1}
\end{figure}

The optical spectrum of a typical PN shows, as we have discussed in Section~\ref{sec:observing_and_modeling_planetary_nebula}, a number of emission lines. An example is shown in Figure~\ref{fig:optical_Tc1}. The brightest are hydrogen and helium recombination lines and oxygen, nitrogen, and sulfur forbidden lines. These elements are the most abundant in PNe, apart from carbon, which has no bright forbidden lines present in the optical wavelength range. The continuum emission of a PN in the optical range is due to the stellar continuum, and if any, the diffuse radiation, which is the continuum emitted by the nebula itself. 
The Balmer ``jump" is a sudden increase in the nebular continuum emission at wavelengths smaller than 3647~$\AA$, due to the photons produced by the recombination of hydrogen to the electronic level $n=2$ in the nebular gas (it can be clearly seen in Fig~\ref{fig:optical_Tc1}).
The photons produced in this free-bound process have an energy that is the sum of the ionization potential for $n=2$ plus the kinetic energy of the recombining electron; the Balmer jump occurs at the limit in energy (or wavelength) where this kinetic energy is zero. Similar "jumps" can be seen at the corresponding limit for each of the hydrogen series and are called Lyman ($n=1$), Paschen ($n=3$), Brackett ($n=4$), Pfund ($n=5$) and Humphreys ($n=6$) - all named after eminent physicists.

\subsubsection{Infrared Emission} \label{sssec:infrared_emission}

The typical PN IR spectrum can be very rich in different types of emission. The low energy tail of the ionizing star continuum in this range is faint and the dust thermal continuum takes over towards higher wavelengths, peaking around $\sim$40~$\mu$m. Many lines and bands from atomic, molecular and solid state species (dust) can be seen overlaid on this continuum. 

Similarly to the optical, the near-IR ($\sim 1-5$~$\mu$m) spectrum shows faint continuum emission overlaid with a number of atomic emission lines. However, it also shows a number of H$_2$ emission lines, the most abundant molecule in PNe and in the Universe. This molecule emits a large number of lines in the near- and mid-IR ($\sim 5-40$~$\mu$m), many of them detected in PNe \citep[e.g.][]{Hora_etal_1999_H2_Spec}. Observations showed that molecular hydrogen is more frequently detected in bipolar PNe. This is known as Gatley's rule \citep{1988ApJ...324..501Z, 1996ApJ...462..777K}. This correlation is likely associated with higher mass progenitors and the presence of a dense equatorial region. It may also be related to the fact that the central stars of these PNe reach higher effective temperatures \citep{Aleman_Gruenwald_2004, Aleman_Gruenwald_2011, 2006MNRAS.368..819P}. Simulations of PNe have demonstrated the significant contribution of H$_2$ emission from partially ionized regions around such hot central stars. Frequently, these H$_2$-rich PNe also show emission from other molecules.

\begin{figure}[h!]
    \centering
    \includegraphics[width=13cm]{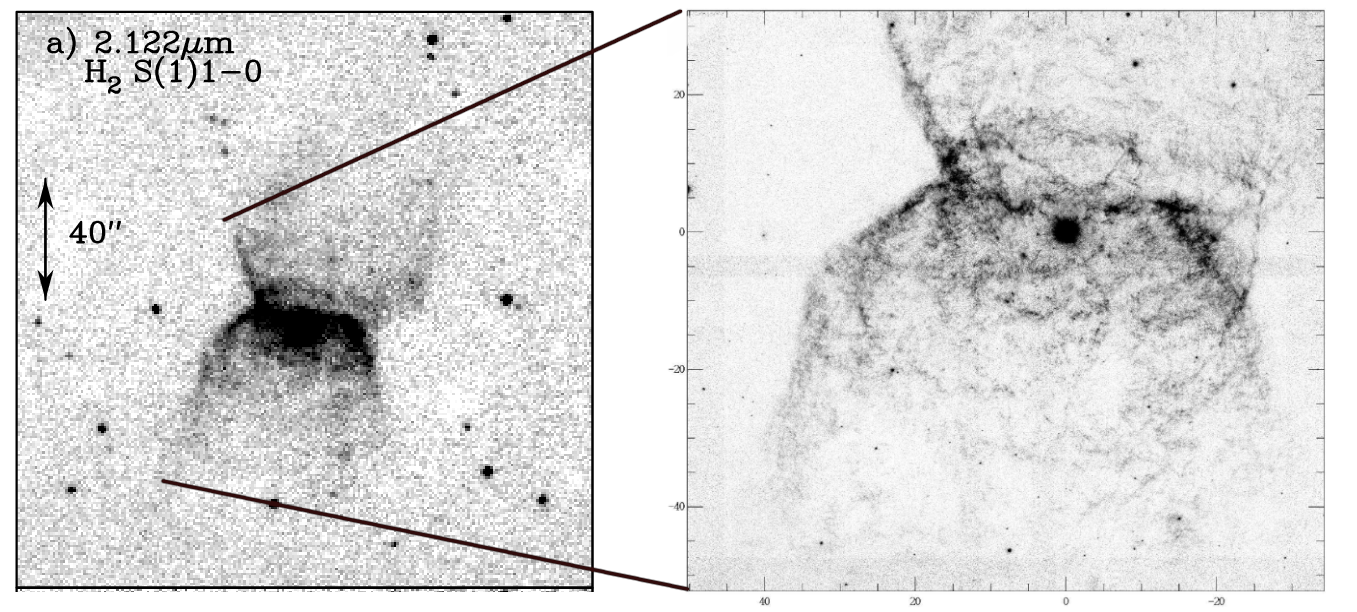}
    \caption{H$_2$ emission images of the PN NGC~2346. Left panel: Low spatial resolution image obtained by \citet[][their figure 2]{Arias_etal_2001}. Right panel: high resolution H$_2$ image obtained by \citet[][their figure 1]{Manchado_etal_2015}, where the fragmentation of the H$_2$ equatorial component becomes evident.} 
    \label{NGC2346}
\end{figure}

High resolution images obtained either from space or with ground-based telescopes show that H$_2$ emission is concentrated in clumps and filaments \citep{Meixner_etal_2005, Speck_etal_2003, Matsuura_etal_2009, Manchado_etal_2015, Akras_etal_2017, Akras_etal_2020, DeMarco_etal_2022, Wesson2024}. This is illustrated in Figure~\ref{NGC2346}, which presents the low spatial resolution (left) and the more recent high-resolution with adaptive optics technique images of the bipolar PN NGC~2346 taken in the H$_2$ 2.122~$\mu$m line (one of the H$_2$ brightest lines in the near IR range). Although often the more intense emission is related to the equatorial regions in bipolar nebulae, H$_2$ is also seen in PN lobes and halos \citep{Kastner_etal_1994_H2_Haloes, Hora_etal_1994, Latter_etal_1995, Guerrero_etal_2000, Arias_etal_2001}.

Another interesting feature of the near-IR spectrum is the presence of emission lines from neutron-capture heavy elements (atomic number $>$ 30). Although first detected in the optical range \citep{Pequignot_Baluteau_1994}, emission lines of the elements selenium and krypton have been detected in the near-IR spectrum of many PNe \citep[][]{Dinerstein_2001,Sterling_2020Galax}. In the near-IR, germanium, bromine, rubidium, cadmium and tellurium were also detected in a few PNe, while in the optical range bromine, krypton, rubidium, xenon were detected in NGC~7027. During the AGB phase, the slow neutron capture process can produce these elements. Their observation in the nebular spectrum is an opportunity to study nucleosynthesis processes in stars of low to intermediate mass.

In the mid-IR, a number of atomic species appear on top of a continuum that is due to dust thermal emission. Typical atomic lines observed in this range includes the forbidden lines emitted by the ions Ne$^{+}$, Ne$^{2+}$, Ne$^{4+}$, Ar$^{+}$, Ar$^{2+}$, Ar$^{4+}$, Na$^{2+}$, S$^{2+}$, S$^{3+}$, O$^{3+}$, Mg$^{4+}$, Cl$^{3+}$, Si$^{+}$, Fe$^{+}$. A few fainter hydrogen recombination lines and H$_2$ lines can also be detected. Deep and high resolution spectra taken with the James Webb Space Telescope (JWST) can show a few hundreds H and H$_2$ lines in the near- and mid-IR \citep{Jones_etal_2023}. 

\begin{figure}[h!]
    \centering
    \includegraphics[width=\textwidth]{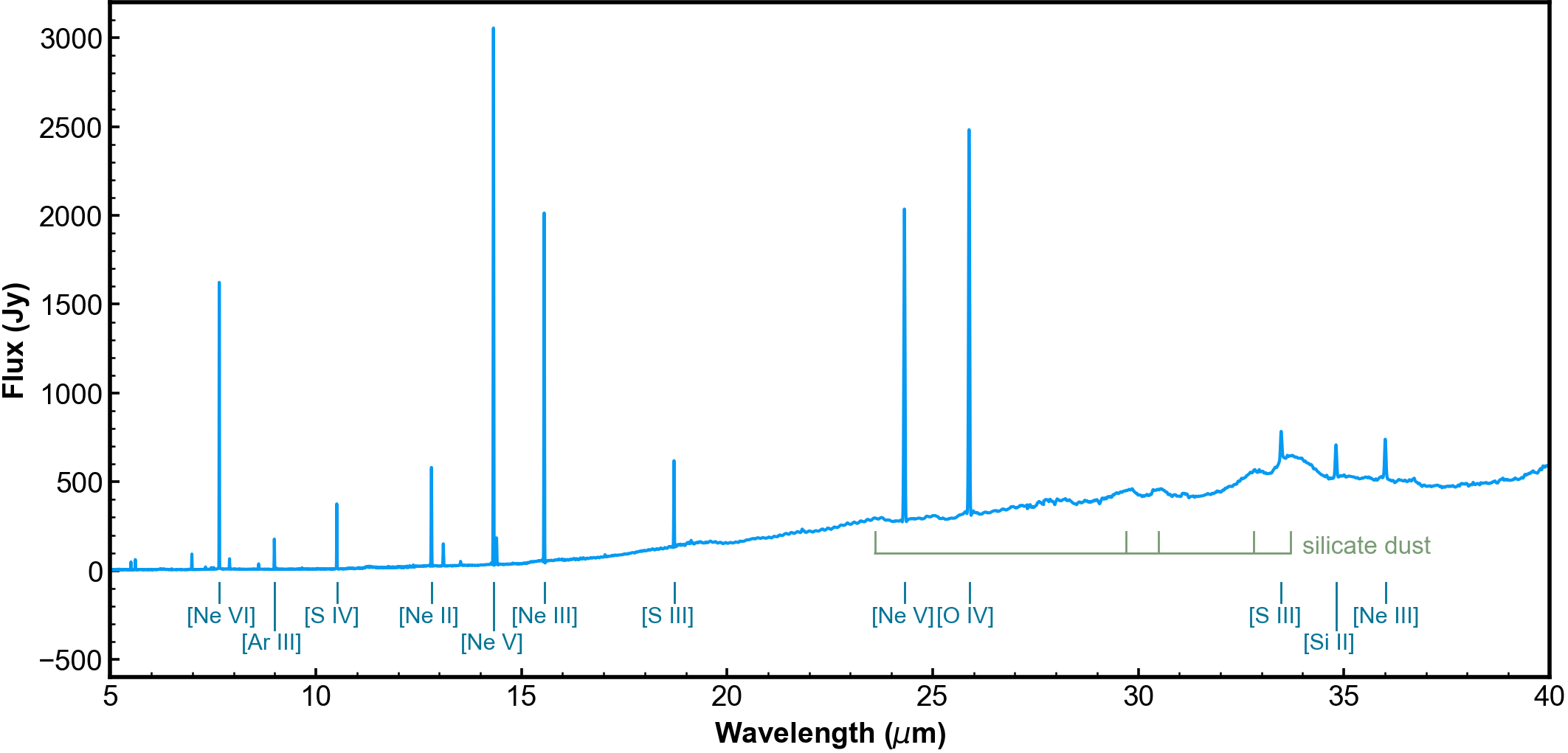}
    \caption{The mid-IR spectrum of the high-excitation PN NGC~6302 taken with the Infrared Space Observatory (ISO). The brightest atomic emission lines are labeled. The spectrum also shows features associated with silicate dust. \citet{Molster_etal_2001_NGC6302} showed that these features are very similar to those of fosterite (Mg$_2$SiO$_4$).} 
    \label{midIR_spec}
\end{figure}

In both the near- and mid-IR spectral ranges, a few bands attributed to vibrations of carbon-carbon and carbon-hydrogen atomic pairs inside molecules may be present. Such bands are attributed to large molecules, as polycyclic aromatic hydrocarbons (PAHs) and its variants \citep{Tielens_2005}. Emission bands from fullerenes (large carbon molecules arranged in a cage structure) were first detected in the PN Tc~1 \citep{Cami_etal_2010_fullerene}.
Solid state features can also be present in the mid- and far-IR spectrum of PNe (Figure~\ref{midIR_spec}). The broad features around 11 (\lq\lq big 11\rq\rq), 20, and 30~$\mu$m have been studied and, although suggestions have been made, their carriers are still under debate \citep{1985ApJ...290L..35G, 2001A&A...378L..41H, 2011ApJ...738..121C, 2017IAUS..323..121S}.

Figure~\ref{midIR_spec} shows an example of a mid-IR spectrum of a PN observed with the Infrared Space Observatory (ISO); NGC~6302 is a high excitation PN that shows lines from highly excited ions as [O~{\sc iv}] at 25.9~$\mu$m, [Ne~{\sc v}] at 14.3 and 24.3~$\mu$m, and [Ne~{\sc vi}] at 7.7~$\mu$m, on top of the dust continuum. Solid state features from silicate grains are also observed. Comparison to spectra obtained in the laboratory for different species, indicate that the set of features are very similar to the spectral features produced by the dust species fosterite \citep[Mg$_2$SiO$_4$;][]{Molster_etal_2001_NGC6302}. 

The far-IR range ($\sim$40-200~$\mu$m) of PN spectra, similarly to the mid-IR, shows atomic and molecular lines on top of a dust thermal continuum. A few very intense lines emitted by atomic species usually stands out: [O~{\sc i}], [O~{\sc iii}], [N~{\sc ii}], [N~{\sc iii}], [C~{\sc ii}]. Such lines are important gas coolers in the region they are emitted. Several important molecular lines have been detected in this range, such as HeH$^+$, CO, CH$^+$, OH, OH$^+$, H$_2$O \citep{2010A&A...518L.144W, 2014A&A...566A..79A,Etxaluze_etal_2014, 2020ApJ...894...37N}.

\subsubsection{Ultraviolet and X-ray Emission} \label{sec:uv_and_xrays_emission}

UV and X-ray radiation is expected in the central regions of PNe, where the gas is fully ionized and its electronic temperature is very high. The central star can also emit in the UV and, if hot enough, in X-rays. In the UV regime, PN spectra are a combination of  stellar and nebular emission. The central star emits a continuum spectrum similar to a blackbody that peaks in the UV due to its high effective temperature. The nebular gas contributes to the continuum emission, as well as emission in recombination and forbidden  lines. The lines of He~{\sc ii} at 1640~\AA, C~{\sc iv} at 1549~\AA, and C~{\sc iii}] at 1909~\AA\ are examples of prominent lines, in particular the semi-forbidden carbon line, as only carbon recombination lines can be observed in the optical spectral range. Other bright carbon forbidden lines are seen in the far IR and submillimeter spectral ranges. The total intensity of UV emission lines can be similar to the total intensity of emission lines in the optical spectral range.

UV emission is absorbed by the Earth's atmosphere. Therefore, the first UV observations of PNe (IC~2149) were performed from space, with the space observatory Orion-2 on board the Soyuz-13 spacecraft in December 1937 \citep{Gurzadian1975}. A year later, spectroscopic data of the PN NGC~7027 was obtained using an Aerobee rocket \citep{Bohlin1975}. UV spectroscopic observations of PNe become more systematic in the 1990s, with the International Ultraviolet Explorer (IUE), the HST and the Far Ultraviolet Spectroscopic Explorer (FUSE). These observations confirmed the presence of emission lines from highly ionized atoms such as neon, verifying the presence of hot central stars ($>$80000~K) at the center of PNe. Furthermore, the detection of highly ionized atoms in the UV that are absent in the visible allows one to constrain better the chemical abundances in PNe.

\begin{figure}[h!]
    \centering
    \includegraphics[width=\textwidth]{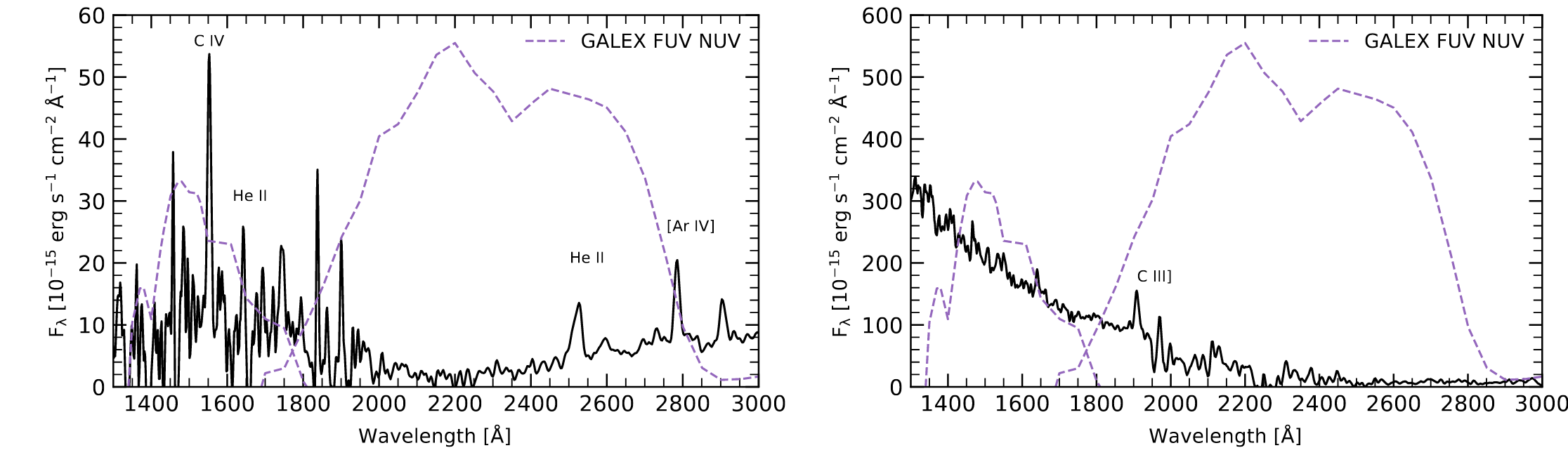}
    \caption{Examples of UV spectra from the International Ultraviolet Explorer (IUE) telescope for two PNe (left panel; NGC~1501, right panel; NGC~3587) with the transmission curves of the two Galaxy Explorer (GALEX) telescope filter profiles (dashed curves) overlaid \citep[adapted from figure 3 of][]{GomezMunoz2023}.}. 
    \label{UVPNe}
\end{figure}

Images of 400 PNe in the far-UV (1344-1786~$\AA$) and near-UV (1771-2831~$\AA$) with the Galaxy Evolution Explorer (GALEX) demonstrated a wide variety in morphologies \citep{Bianchi2018}. Some PNe have shown significantly different radial profiles in the two bands, indicating complex ionization structures. The richness of UV emission line spectra of PNe is shown in Figure~\ref{UVPNe}, for two PNe. There we can also notice that the continuum emission from the hot central star can be quite variable. For NGC~3587, the presence of the hot star is clear in the continuum emission, rising for shorter wavelengths. A (far UV -- near UV) color smaller than zero, has in fact been widely used to identity hot compact sources in the GALEX catalogs, including single hot white dwarfs ($>$20\,000~K), binary PN central stars, symbiotic stars, or cataclysmic variables \citep[e.g.,][]{Bianchi2011,Akras2023,GomezMunoz2023}.

\begin{figure}[h!]
    \centering
    \includegraphics[width=16cm]{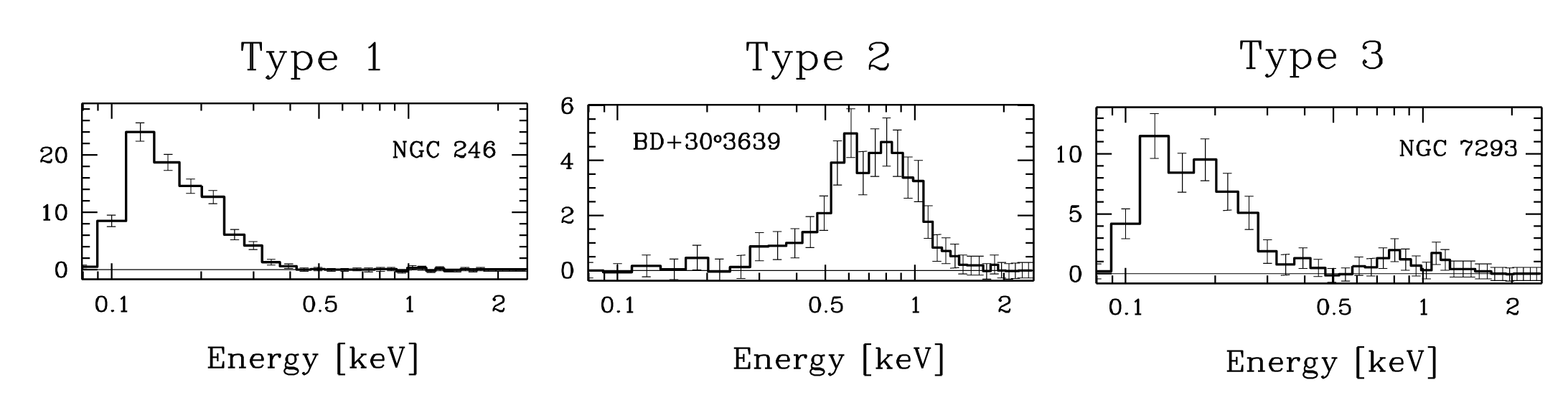}
    \caption{Examples of the three types of PN X-ray spectrum, based on ROSAT observations. Adapted from figure 3 of \citet{Guerrero2000}.} 
    \label{typesxrayPNe}
\end{figure}

PN X-ray emission was discovered in the 80s and 90s with observations carried out by the Einstein Observatory, the European X-ray Observatory Satellite (EXOSAT) and the Roentgen Satellite (ROSAT). The first detection of X-ray emission was made in the PN NGC~1360, with observations from EXOSAT by \cite{deKorte1985}. The ROSAT observations, with much better sensitivity and spatial resolution,  resulted in the detection of X-ray emission from 63 PNe. 
Three types of PNe X-ray spectrum were recognized based on these data \citep{Guerrero2000}: Type~1 sources, with soft X-ray emission ($<$0.5~keV) peaking at 0.1–0.2~keV, corresponding to a T$\sim$(1-2)$\times$10$^5$~K blackbody spectrum; Type~2 sources, with harder X-ray emission peaking at 0.5~keV, which corresponds to a T$\sim$1$\times$10$^6$~K optically-thin gas emission; and Type~3 sources, with a combination of Type~1 and Type~2 spectra. Figure~\ref{typesxrayPNe} shows X-ray ROSAT spectra of each PN X-ray type.

Moreover, the PN X-ray emission can be either point-like or diffuse. The point-like sources in PNe are associated with their central star. The hot central star ($>$100\,000~K) can account for the X-ray emission of photons with energies $<$0.3~keV. However, some point-like sources also show high luminosities and energies $>$0.5~keV, and these cannot be explained by any photospheric temperature, nor any other process associated with their white dwarf central stars. In these cases, the X-ray emission is likely linked with the shock interaction of their stellar winds in a close binary system or with the hot corona of their main sequence-type companions \citep[][and references therein]{Montez_etal_2015}.

The origin of the diffuse, as opposed to point-like, X-ray emission is instead associated  with the interaction of the fast stellar wind (1000-3000~km~s$^{-1}$) from the post-AGB phase, with the slower winds from the preceding AGB phase. This interaction results in the production of a reverse shock, which propagates inwards and interacts with the stellar wind. This process results in the formation of a shock-heated, overly pressurized bubble, with temperatures up to 10$^6$~K, capable of emitting soft X-rays. The boundary between this hot bubble and the undisturbed, ionized, post-AGB gas is defined by the dense rim nebular structure, discussed in Section~\ref{sec:nebularstructres}. 

Young PNe ($<$5000 years) with high electron densities ($>$1000 cm$^{-3}$) are the sources of diffuse X-ray emission. Moreover, more massive central stars produce brighter and hotter X-ray bubbles, due to the significantly large amount of mechanical energy contributed by their strong stellar winds in a very short time and space \citep{Ruiz2013}.  

\begin{figure}[h!]
    \centering
    \includegraphics[width=15cm]{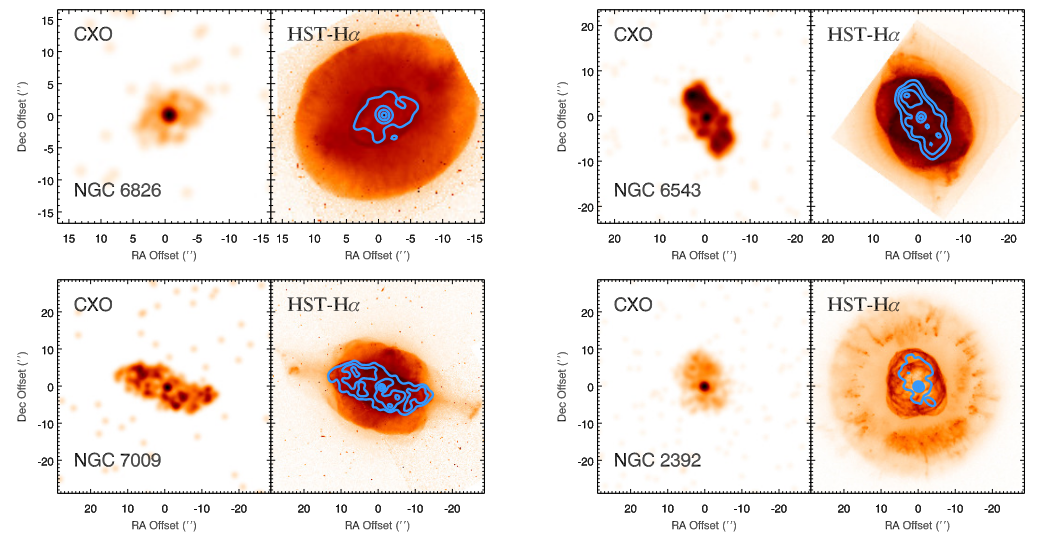}
    \caption{Diffuse X-ray emission detection from {\it Chandra} in 4 well-studied PNe (NGC~6826, NGC~6543, NGC~7009 and NGC~2392). In each panel, the X-ray emission is shown on the left, while on the right panel the X-ray emission contours are overlaid on the H$\alpha$ images. Figures from figures 2 and 3 of \cite{Kastner2012}.}     
    \label{diffusexrayPNe}
\end{figure}

The Chandra Space Telescope ({\it Chandra}), X-ray observations indicate a high detection rate of diffuse and point like X-ray sources for close PNe ($<$1.5~Kpc; 70 percent according to \cite{Kastner2012} and 54 percent according to \cite{Freeman2014}).
In Figure~\ref{diffusexrayPNe}, we display as examples the {\it Chandra} X-ray images and their contours overlaid on optical images for four well studies PNe. These images illustrate the presence of hot bubbles at the center of these nebulae. In all cases, the bright rim appears to be the outer border of these hot bubbles.

\subsubsection{Submillimeter to Radio Emission} \label{sssec:radio_emission}

\begin{figure}[h!]
    \centering
    \includegraphics[width=12cm]{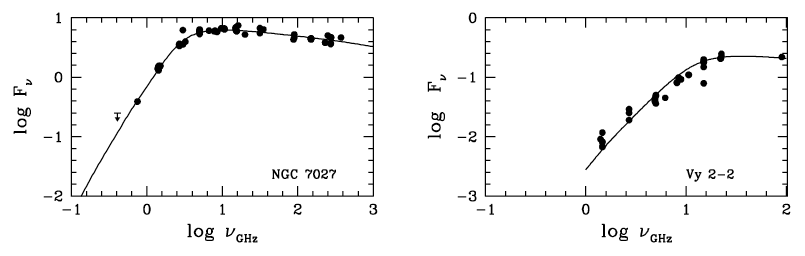}
    \caption{Radio spectra of two PNe. Dots represent observations, while the curves are obtained from photoionization models. Figure adapted from figure 5 of \citet{2007A&A...461.1019G}.} 
    \label{fig:radio}
\end{figure}

PNe are not usually strong submillimeter or radio emitters. The low-energy tail of the thermal, blackbody dust continuum reaches the submillimeter region of the spectrum ($\lambda \sim$ 0.3-1~mm, which corresponds to 30~GHz to 1~THz in frequency), but it is faint compared to the IR emission at shorter wavelengths. At even longer wavelengths, i.e., lower energies/frequency ($\lambda >$ 1~mm), the PN continuum is dominated by emission from the ionized gas, known as free-free emission (\textit{bremsstrahlung}), photons produced when free electrons are slowed down by positively charged hydrogen and helium nuclei. As the nebula is optically thin for a large range of frequencies in the radio spectral range, this mechanism helps to cool the gas. Examples of PNe radio spectra are shown in Figure~\ref{fig:radio} \citep{2007A&A...461.1019G}.

As in other spectral ranges, atomic and molecular lines produced by the nebular gas are seen overlaid to the ionized gas continuum. Submillimeter and radio recombination lines of hydrogen, as well as forbidden lines from atomic carbon have been detected  \citep{2014A&A...566A..79A, Etxaluze_etal_2014}.
Several molecules have been thus identified (CO, OH, OH$^+$, etc.). Of these CO is very plentiful. It is believed to survive in the outskirts of PN envelopes and in dense regions such as in torus-like and ring structures around the central stars, where it is shielded from the UV radiation of the star \citep[e.g.,][]{Huggins_etal_2005,2018A&A...618A..91G}. 
Recent observations of lines of more complex molecules have been detected in the submillimeter revealing a previously underappreciated chemical richness in PNe, which complements what we knew of its composition, physical conditions, evolution and structure \citep[e.g.,][]{2014A&A...566A..79A,2014ApJ...794L..27E,2016ApJ...817..175S,2018IAUS..332..218S, 2019ApJ...881L..38S, 2019A&A...625A.101B}. 

OH maser line emission have also been detected in a small number of PNe \citep[OHPNe;][]{Zijlstra_etal_1989}. Masers, like lasers (the only difference is that the radiation is in the microwave wavelength instead of in the optical) are usually faint but when detected they provide precise information on the gas conditions, velocity and magnetic fields. These maser emitting PNe are usually newly formed. 

\subsection{Distances}\label{sec:distances}

Accurate distances are needed in order to determine PNe's intrinsic properties, such as the physical dimensions, the luminosity, and the overall energies involved in the formation and evolution of these objects. Measuring distances to galactic objects is notoriously challenging. The values obtained with a variety of methods are often discordant and inaccurate. A comprehensive overview of the distance determination methods used for PNe is given by \cite{Frew2016}. 

The distance measuring methods can be categorized into two main groups: statistical methods and individual methods. The most direct distance measurements to individual PNe is by trigonometric parallax. It is based on the apparent displacement of a star in the sky, compared to other more distant, background stars, due to the change in the observer's position as the Earth moves around the Sun. The {\it Hipparcos} satellite measured trigonometric parallaxes and distances for millions of stars, including some nearby central stars of PNe. With the more advanced {\it Gaia} satellite, aiming to measure accurate trigonometric parallaxes for over a billion stars, many more central stars were surveyed. The first cross-match of known central stars with the Gaia Data Release 2 (DR2) yielded a sample of 655 sources, of which 430 have parallax measurements and small uncertainties \citep[e.g.,][]{GonzalezSantamaria2019}. 

Another individual method is the ``expansion parallax" method. If one can measure the expansion of the nebula on the plane of the sky over a period of time (usually a decade or more), jointly with the physical speed of the expansion, one can derive the distance using the equation $D_\textrm{par} = 211 \times~ V_\textrm{Doppler}$/($d\theta$/dt), where $V_\textrm{Doppler}$ is the velocity measured using the Doppler effect, $d\theta$ is the apparent expansion on the plane of the sky and $dt$ is the time period over which the expansion took place. As always there are caveats to all methods, in this case the fact that the velocity measured through the Doppler effect is that along the line of sight and an assumption is made that it applies to the gas that we see moving on the plane of the sky, hence perpendicular to the light of sight, this assumption is not always correct \citep{Mellema2004,Schonberner2005}.

Statistical methods use instead samples of PNe with specific assumptions regarding the average nebular structure and properties. \cite{Frew2016} compiled a catalog of statistical distances for nearly 1100 Galactic PNe, using the assumption that the total PN brightness in the H$\alpha$ line (Balmer line, transition between levels 3 and 2 in hydrogen) is anti-correlated with the radius (that is, the PN expands and fades). If one can calibrate this correlation for a few objects for which distances are known from an alternative method (such as from the trigonometric parallax method), one can then use the measured H$\alpha$ brightness of a PN and use the relation to obtain its physical size, which together with the measured apparent size on the sky can give us the distance to the object.  However, one must remember that even if the relation between the brightness and the size is robust, there may be substantial scatter so that this method is best applied to groups of objects for which average properties are needed, rather than to individual ones. A few years later, \cite{Ali2022} used a similar statistical distance scale based on the linear relationship between radio surface brightness temperature, nebular radius, and more reliable {\it Gaia} parallax measurements as calibration.
Utilizing these results, the surface brightness versus radius relationship was recalibrated \citep{Stanghellini2020}. 
A subsequent effort by \citet{GonzalezSantamaria2021} presented a catalog of 1725 PNe with reliable parallaxes provided by {\it Gaia} in the Gaia Early Data Release 3.

\subsection{Modeling Planetary Nebulae} \label{ssec:modeling_planetary_nebulae}

\begin{figure}[h!]
    \centering
    \includegraphics[width=15cm]{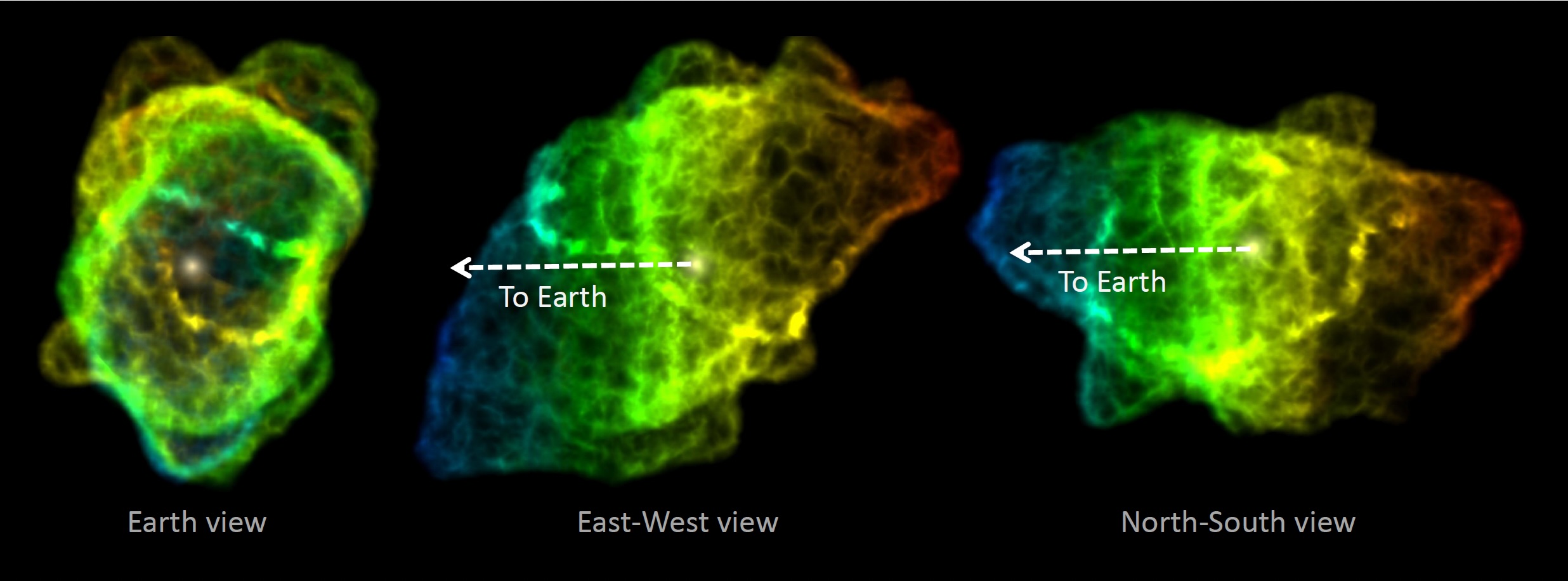}
    \caption{Three views of the Southern Ring nebula from morpho-kinematic models (figure 3 of \citet{DeMarco_etal_2022}) using the morpho-kinematic modeling code {\sc shape} \citep{Steffen2006}.}     
    \label{fig:SoutherRing}
\end{figure}

We model PNe to test whether we understand the main physical mechanisms that play a role in generating the nebula we observe. Ideally we would like to have a comprehensive model that, {\it ab initio} generates a synthetic image of the nebula (along with spectra of the star and nebula) that reproduces the observations. In practice, this would be too complex to be computed in feasible times and we therefore tend to model aspects of the star and nebula in isolation. This said,
modern codes are more and more complete and complex. Computational codes are available to model PN from zero to three dimensions. Accurate atomic and molecular data valid for the PN physical conditions are essential for these models and are becoming increasingly available as more laboratory studies, in part motivated by astrophysics, are being published. Finally, faster computers allow modeling codes to increase the number and complexity of the included physical mechanisms.

Photoionization models are numerical simulations constructed with the aim to determine the nebular ionization structure, the physical conditions and the radiative emission from the PN. There are many physical and chemical processes occurring simultaneously. They depend on the physical conditions which, in turn, depend on the processes. Iterative calculations of complex and numerous equations are then necessary for a detailed and precise model, leading to the development of photoionization codes \citep[e.g.,][]{Ferland_etal_1995}. These models were originally designed to simulate only the ionized regions of PNe, but are now expanded to include the processes necessary to also simulate the neutral and molecular regions, affording us a complete view of the PN structure. Full, three-dimensional photoionization models are computationally expensive, but can provide a very detailed view of the PN \citep[e.g.,][]{Monteiro_etal_2004}. A simplified, but still powerful option is a pseudo-3D model \citep[e.g.,][]{2016A&A...585A..69G,Aleman_etal_2019_Tc1}. Such models are constructed from the combination of multiple runs of one-dimensional models. The limitations are second order effects, thus providing very good results.

Morpho-kinematic models aim primarily to reproduce the 3D shape of the PN. These models use, as input, spectroscopic information of different parts of the PN. Each part of the PN projected on the plane of the sky has a line-of-sight motion that is detected by the spectrum. An educated guess is then made on how the PN gas is moving with respect to the central star, so that the Doppler information from the spectrum can be translated into a distance from the central star. This gives the location of different parts of the PN gas along the line of sight. The location of gas on the plane of the sky, as measured directly from images, gives the distance to the central star in the perpendicular direction. A 3D image is then reconstructed \citep[Figure~\ref{fig:SoutherRing}; e.g.,][]{2011ITVCG..17..454S,Akras2016,DeMarco_etal_2022}. A morpho-kinematic reconstruction can be used as the initial matter distribution to develop a photoionization model. This powerful combination allow us to faster obtain more accurate models for individual objects.

\begin{figure}[h!]
    \centering
    \includegraphics[width=15cm]{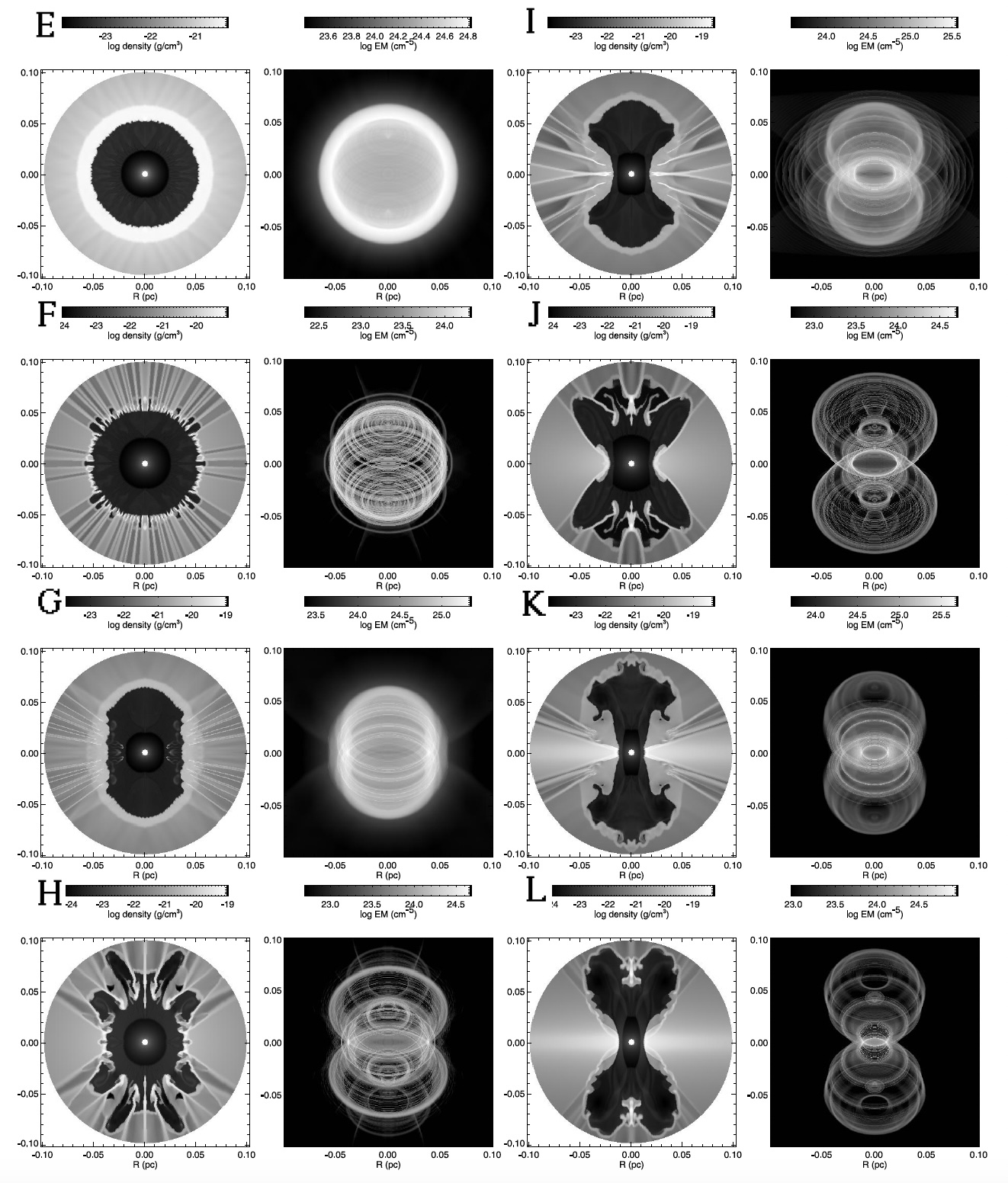}
    \caption{Modelled density slices (first and third columns) of a number of PNe for which rotation and the amount of ionizing stellar flux is varied. 3D images of the emission (second and fourth columns) for which the viewing angle is 45~degrees. (Figure 2 of \citet{GarciaSegura1999}).}     
    \label{fig:garcia-segura}
\end{figure}

Hydrodynamic or magnetohydrodynamic (including magnetic fields) simulations have also been used to study the formation and evolution of PN structures (lobes, tori, rings, or jets). The key idea of those  simulations is that stellar rotation and the action of magnetic fields promotes stellar mass-loss with an equatorial concentration. The star that ejects the nebula would then be left in the middle of a giant doughnut-like structure. When the central star ejects the fast, tenuous wind later on, it would blow into this torus structure, resulting in lobes inflating in a direction perpendicular to the torus and thus forming a bipolar nebula. By modulating the rotation and magnetic fields (as well as other parameters) a number of structures can form from very collimated to elliptical. With this relatively simple mechanism, these models appeared to explain the overarching shapes of the majority of PNe \citep[e.g.,][see Figure~\ref{fig:garcia-segura}]{Mellema1994,GarciaSegura1999}. Some of these results were challenged in the last 20 years as we explain in Section~5.

\begin{figure}[h!]
    \centering
    \includegraphics[width=13cm]{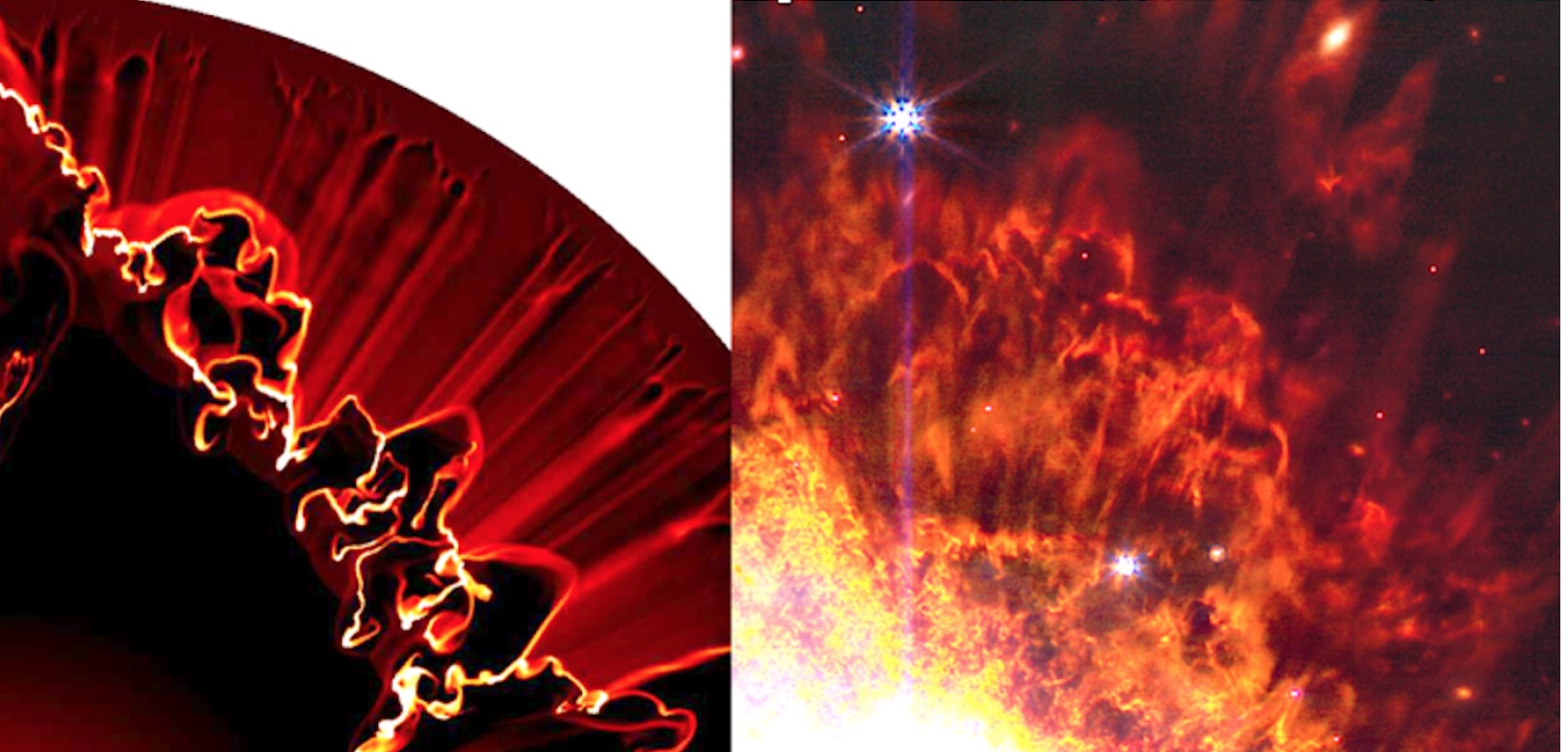}
    \caption{Left panel: a hydrodynamic simulation showing the formation of nebular structures external to the main ionized region of the Southern Ring nebula, NGC~3132. Right panel: a zoom in of the JWST observation of the Southern Ring nebula. Adapted from figure 4 of \citet{DeMarco_etal_2022}.}     
    \label{fig:fingers}
\end{figure}

Smaller scale structures such as knots, filaments, jets and rings are explained by a number of smaller scales physical mechanisms such as shocks, illumination, dust, etc. with variable level of success (see Figure~\ref{fig:fingers}, where the models of \citealt{Villaver2002} were adopted by \citealt{DeMarco_etal_2022}). In a minority of cases, the presence of structures must denote the action of a companion, interacting with the progenitor giant star during the formation of the PN. For example very collimated jets may indicate an accreting companion which launches a jet from an accretion disk.

\section{Observing and Modeling the Central Stars}

\label{sec:observing_and_modeling_the_central_star}

\begin{figure}[h!]
    \centering
    \includegraphics[width=13cm]{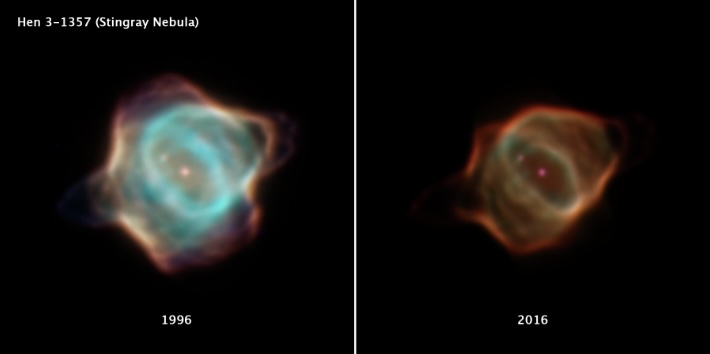}
    \caption{Dramatic fading of the nebula known as the Stingray (He~3-1357), caught between 1996 (left) and 2016 (right) by the Hubble Space Telescope. This is due to the star evolving and fading, something that is mirrored in the fading of the PN. Credit: NASA, ESA, B. Balick (University of Washington), M. Guerrero (Instituto de Astrofísica de Andalucía), and G. Ramos-Larios (Universidad de Guadalajara).}     
    \label{fig:stingray}
\end{figure}

Central stars of PNe are caught in an evolutionary phase between the large and cool AGB stars (the size of Earth's orbit) and the tiny but hot white dwarfs (the size of Earth). The change that takes place to stars in this phase is not only extreme, but can also be extremely fast (see Figure~\ref{fig:stingray}), and can be observable over decades. 

AGB stars have by and large, carbon and oxygen cores, and large hydrogen-rich envelopes. A smaller, originally more massive, subset can have oxygen and neon cores. Between core and envelope there are, typically, two thin burning shells. The inner helium shell still burning as the star expands away from the horizontal branch (when it was burning helium in the core) onto the early AGB, and an outer hydrogen shell surrounded by a large, convective envelope. AGB stars grow in radius and core mass as they increase in luminosity while also loosing envelope mass until at some point the mass-loss rate increases, likely due to pulsations and the formation of dust, and almost the entire envelope is ejected over a small period of time. The envelope that is left has such low mass and density that there is no solution to hydrostatic equilibrium and the photospheric radius decreases, with a subsequent increase of the effective temperature. This explains the left-ward shift on the HR diagram (Fig.~\ref{PN_evol_new}, brown line).

\subsection{Observations of Central Stars of PNe and Atmospheric Models}\label{sec:observations_cspn}

Because of the uncertain distances plaguing Galactic PNe before the arrival of the {\it Gaia} space telescope, observations of central stars of PNe in the Magellanic Cloud galaxies were the first to provide accurate parameters. At the known distance of the Magellanic Clouds, the central star brightness can be corrected for distance and the true brightness determined. Then, via the determination of the surface gravity and surface temperature from a spectroscopic analysis, the mass can be obtained. This exercise revealed that central star masses appear to be in a relatively narrow range 0.65$\pm$0.07~$M_\odot$ \citep{Villaver2007}. This has allowed us to discover that there is a core mass-luminosity relation \citep[e.g.,][]{Vassiliadis1994} and even assess the evolutionary speed (that translates in a photospheric heating rate) between the AGB giant cool phase, to the white dwarf, hot end-point \citep[e.g.,][]{Stecher1982,Villaver2003,Villaver2007}. These observations are critical to assess the quality of the stellar interior models. Besides a handful of central stars located in clusters \citep[like PHR 1315-6555 in the open cluster Al~1;][]{Fragkou2019}, the only other accurate characterization of their parameters has been enabled by the {\it Gaia} telescope, that has afforded a new generation of parameters of Galactic central stars.

The spectrum of central stars of PNe exhibit a number of different typologies. Regular, hydrogen rich PN central stars are a majority. The hotter types are indistinguishable from hydrogen-rich (known as DA) white dwarfs. About a quarter of all central stars are hydrogen-poor, with some presenting emission lines due to strong winds and high mass-loss rates. These are often referred to as Wolf-Rayet, or [WR] central stars, because in some cases their spectra are indistinguishable from those of massive Wolf-Rayet stars (the addition of the square bracket is to distinguish low mass central stars of PNe from massive stars). They are mostly present in two of the three varieties typical of massive Wolf-Rayet stars, the carbon and the oxygen rich ones ([WC] and [WO], respectively). More recently, a handful of the third type, the nitrogen type, or [WN] central stars, have also been detected, where nitrogen (and some hydrogen) lines dominate. Abundances of [WR] central stars, when carefully measured, show approximately 50\% helium and 50\% carbon (He : C : O = 0.30-0.85 : 0.15-0.60 : 0.02-0.20; \citealt{Werner2006}).

Another type of hydrogen-deficient central stars are the so-called PG1159 stars. These have similar abundances to the [WR] central stars, but have much weaker winds as demonstrated by the weaker emission lines in their spectra. A fraction of these stars do not present a visible PN. [WR] and PG1159 central stars likely evolve in a scenario where a last helium shell thermal pulse takes place after the star leaves the AGB \citep{Bloecker1997,Bloecker2001,Herwig2001}. Depending on the exact timing of this late thermal pulse, the abundances predicted vary, but they would ultimately reflect the inter-shell values (this is the region between the hydrogen and helium burning shells). 

Finally, there is another class of hydrogen-deficient central stars, which have dramatically different abundances: they are mostly made of helium, and are related to the so called ``extreme helium stars''. While the extreme helium stars with no PN may be the descendant of mergers, some of the central stars with the same characteristics may derive from more complex scenarios such as the merger of a white dwarf with the core of an AGB star inside a common envelope \citep{Reindl2014}.

To complete the lineup of central stars of PNe, there are, at last, some that seem to have descended from RGB stars. In other words, they have not yet become  AGB stars (and may do so in the future). They make it into the central star of PN list because the nebulae that surround them are very similar to PNe. We may elect to exclude these objects from our catalog of central stars, but they retain an associated interest because somehow they have ejected and ionized the nebular material, clearly doing so in some related mechanism.

\subsection{Modeling the Interior}\label{sec:modeling_the_interior}

Early computational models of stars evolving through the AGB phase were those of \citet{Paczynski1970,Paczynski1970b, Paczynski1970c}. The complexity derived from the need to follow critically the two burning shells, which have small masses and tiny physical extent, necessitating short time steps. Additionally, somewhere along the AGB, it was known that the star loses from about half to 90~\% of its mass in relatively short times due to an unidentified mechanism that would need to be reproduced in the models.

The next stint of progress in modeling the structure of PN central stars came in the 90s with major efforts such as those of \citet{Bloecker_Schoenberner_1990},\citet{Vassiliadis1994}, \citet{Bloecker1995a}, \citet{Bloecker1995b}, and \citet{Schoenberner_Bloecker_1996}. These papers determined the way we think of post-AGB stars for at least three decades. They determined that there is a relation between core-mass and luminosity; they explained why PNe are either carbon or oxygen rich; they modelled the different speeds at which each star heats from AGB to white dwarf temperatures, including the time it takes the star to heat up sufficiently ($\sim$25\,000~K), to ionize the PN gas (called the ``transition time").

The existence of hydrogen deficient central stars, such as [WR] and PG1159 central stars alerted Astronomers to the need for a mechanism, acting about a quarter of the times, that allowed a star to consume or eject the last remaining hydrogen-rich envelope at the end of their AGB evolution. Another observational indication of the need for a hydrogen-elimination mechanism was the discovery of PNe in which an outer hydrogen rich gas shell surrounds an inner, clumpy, hydrogen poor structure, as if a second ejection had taken place at a later time after the first, AGB PN ejection. 

Early models to explain hydrogen deficiency in the post-AGB evolutionary tracks go back to \citet{Fujimoto1977} and \citet{Iben1995}. They realized that once in a while an AGB thermal pulse happens {\it after} the star leaves the AGB. This is because the process regulating the frequency and number of helium shell thermal pulses is (mostly) independent of the process that regulates the mass loss which leads to the loss of almost the entire envelope and the departure from the AGB (the start of the post-AGB  evolution). This means that the AGB departure can take place at any moment between two successive shell pulses. When departure happens sufficiently close the next pulse, that pulse takes place after the AGB, but before the cessation of nuclear burning (before the star becomes a white dwarf).  In particular, there are three specific situations when a ``late" thermal pulse can take place, named the AGB {\it final} thermal pulse (the last pulse happens right at the end of AGB evolution), the {\it late} thermal pulse (the last pulse happens in the immediate post-AGB time) and the {\it very late} thermal pulse (the pulse happens  at the end of nuclear fusion near the white dwarf cooling track), depending on when in the post-AGB phase the last pulse happens \citep{Herwig2001}. Each results in a particular amount of left-over hydrogen, or in no hydrogen at all. Many questions remain regarding hydrogen deficient central stars that may only be answered when  including companions interacting with the AGB star before it has fully ejected its envelope or the inclusion of 3D processes in the calculation of the structure \citep[e.g.,][]{Herwig2014}.

\begin{figure}[h!]
    \centering
    \includegraphics[width=13cm]{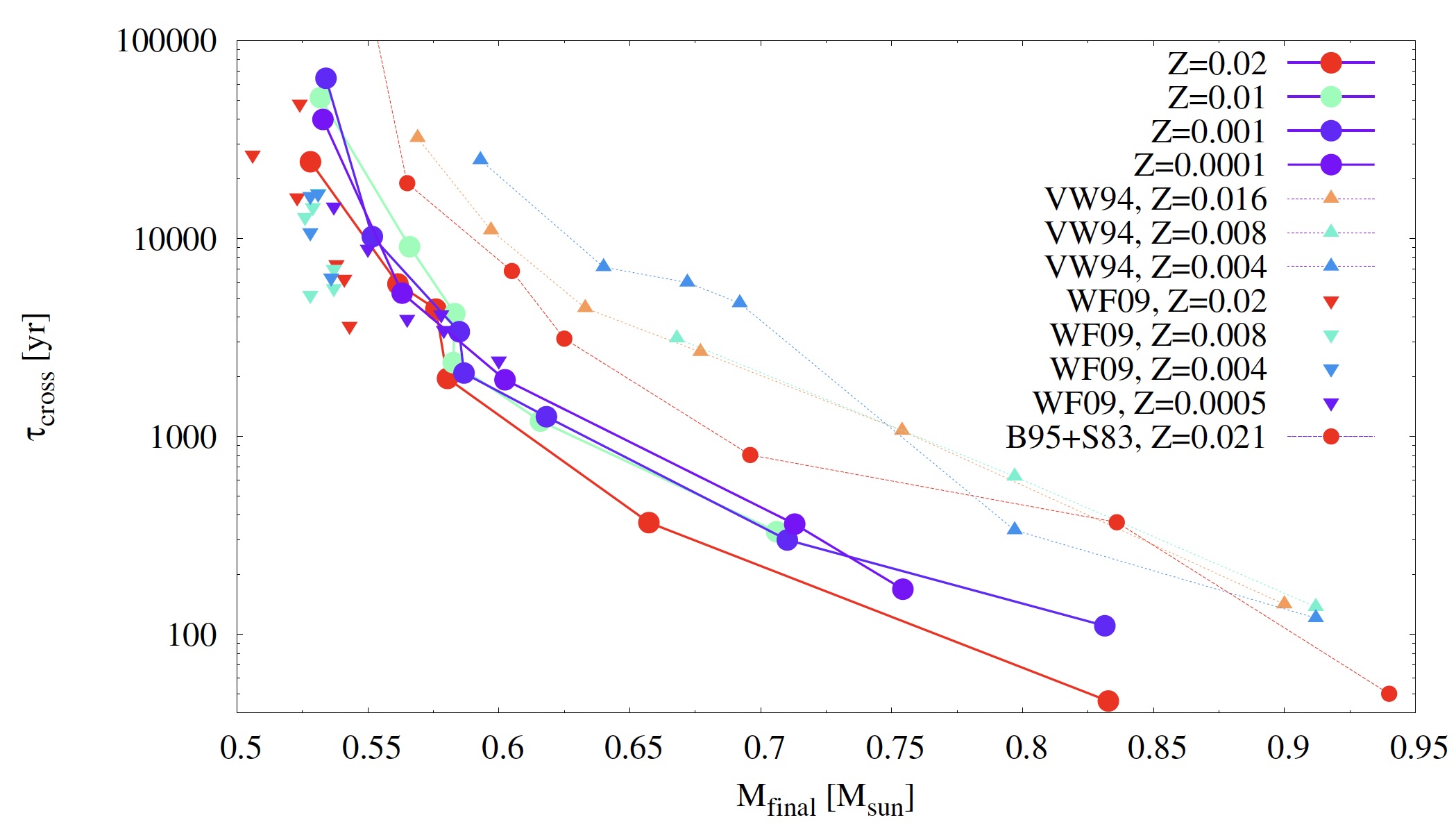}
    \caption{Predicted transition times ($\tau_{\rm cross}$; the time for a post-AGB star to warm up sufficiently to ionize its PN) as a function of central star mass ($M_{\rm final}$), using the models of \citet[][; large circular symbols where the predictions are for different metallicity values, Z]{Miller_Bertolami_2016}. The other models are for different, earlier studies, generally predicting longer times. (Figure 9 of \citet{Miller_Bertolami_2016}).}     
    \label{fig:miller}
\end{figure}

The next stint of progress in modeling central stars' interiors happened with the new models of \citet{Miller_Bertolami_2016}. They predicted much shorter transition times - the time it takes a post-AGB star change between a cool and large AGB star to a star hot enough to ionize the nebula (approximately 25\,000~K; see Figure~\ref{fig:miller}). This meant that lower mass stars than previously thought could reach PN ionization temperatures before the PN gas disperses, indicating more stars could make PNe.

A fundamental problem with 1D stellar interior models is that some processes are inherently three-dimensional. An example is convection, which in 1D  is parameterised by the famous ``mixing length'' theory \citep{Boehm-Vitense1958,Joyce2023}, and calibrated via some of the observable consequences of convection. New, 3D models are however starting to be computed \citep[e.g.,][]{Andrassy2022,Blouin2024}, limited fundamentally by the need for high resolution and the inclusion of many diverse physical processes, which lead to extremely long computational times even on the fastest supercomputers.

\section{Binary Stars and Planetary Nebulae}
\label{sec:binary_stars_and_planetary_nebulae}

Until approximately the mid 2000s the predominant idea to explain the diverse shapes of PNe was, among others, that of \citet{GarciaSegura1999}, who explained how a combination of {\it stellar rotation and magnetic fields} during the mass-losing AGB phase can lead to most morphologies (Section~\ref{ssec:modeling_planetary_nebulae} and Figure~\ref{fig:garcia-segura}). A minority of morphologies such as jets (Figure~\ref{fig:images_PNe}, bottom right), which are occasionally observed in PNe, may be only explained with the action of a companion orbiting the AGB star that, by interacting with the AGB mass loss process accretes some gas and powers jets. At that time the only clear advocate of the importance of these binary interactions not just rarely, but commonly in PN shaping was Noam Soker \citep{Soker1997}. 

In 2006 two papers, \citet{Nordhaus2006} and \citet{Soker2006}, were published that explained how magnetic fields in AGB stars cannot be sustained throughout the mass-losing phase, because their presence would slow down the stellar rotation, the very mechanism that generates the fields. The mechanism is like that of an ice skater that extends her arms and slows down: the rotation-generated magnetic field lines are like the arms of the skater; they anchor the extended AGB wind to the stellar surface, effectively slowing down the rotation, which in turn shuts off the magnetic field in a negative feedback loop \citep[see also][]{GarciaSegura2014}. This would support the conclusion that only spherical and possibly mildly elliptical PNe can derive from single stars \citep{Soker1997}. Other PN morphologies would need a companion to the AGB star to resupply the angular momentum through an interaction. This angular momentum can help retain the rotation, magnetic field, or directly help shape the nebula into morphologies such as bipolar.

If only spherical and mildly elliptical PNe can be formed via a single star channel, we would have to conclude that most PNe come from binaries, since the fraction of spherical and mildly elliptical morphologies is less than a third of the population \citep{Parker2006}. This in turn begs the question of whether there are enough binary interactions on the AGB to justify the observed population of non-spherical PNe. This question was answered by \citet{Moe2006} and \citet{Moe2012} who, using a population synthesis technique, predicted that the number of PNe in the Galaxy from single and binary stars should be 6 times larger than the observed population\footnote{It should be pointed out that while the observed PN population is not complete, as it is a brightness-limited sample, more surveys that find ever fainter objects, are not adding scores of objects to the group, pointing to the fact that the observed population is not going to grow by a large number of objects \citep{Moe2006}.}. On the other hand, if only binary interactions make a PN, the number of predicted PNe is only slightly lower than the observed population, pointing to the possibility that binary interactions may indeed be able to explain most PNe.

It was also realized that PN surveys able to detect fainter objects found a disproportionate number of {\it spherical} PNe, indicating that, if we make the logical assumption that spherical PNe come from single stars, then single stars make low brightness PNe. This would explain why the expected large number of PNe from single stars is not observed. This in turn pointed to the fact that it is possible that the brighter PN population has a higher incidence of close binaries than the progenitor population of all 1-8~$M_\odot$ stars.

The next logical question is that if PNe come from binary interactions we should find binaries in the middle of PNe - except maybe for those objects that have merged during the interaction. On the observational front it was already known that at least 10-15\% of all central stars have companions in orbits with period of 1-3 days, detected via light variability, where the variability is due to eclipses, stellar distortion or irradiation heating \citep{Bond2000}. While this limit is only slightly larger today \citep[$\sim$20\%;][]{Miszalski2009,Jacoby2021}, what is clear is that there are likely more binaries with periods as long as multiple years, not easily detected with the variability technique, because the effects that cause the variability decrease with larger orbital separation \citep{DeMarco2008}. The difficulty of finding wider binaries, in particular, makes a direct and definitive measure of the binary fraction of central stars of PN impossible. Some wider binaries have been found in PNe such as the two detected by \citet{Jones2017}. There is also an indication that more binaries (with an unknown orbital period) are there \citep[e.g.,][]{DeMarco2013,Douchin2015}.

The number of hydrodynamic simulations that model a binary interaction and the subsequent PN formation are very few and limited. \citet[][and following papers]{GarciaSegura2018}, \citet{Frank2018} and \citet{Zou2020} used 3D hydrodynamics calculation of the common envelope interaction to then model the formation of a PN from the interaction of a post-AGB fast wind with the common envelope ejecta. They could only demonstrate that the nebulae are bipolar and can suffer substantial asymmetries due to the inherently 3D nature of the common envelope ejection. Another attempt was by \citet{Ondratschek2022} who contrary to the previous simulations used a magneto-hydrodynamics code to model a common envelope interaction, showing that the magnetic field amplification leads to strong bipolar outflows, reminiscent of young bipolar PN. However, they did not model the second phase of fast wind ejection into the slow, AGB (and binary induced) ejecta, nor did they model the ionization phase, the two phases which  create the PN as we know it.

\begin{figure}[h!]
    \centering
    \includegraphics[width=7.2cm]{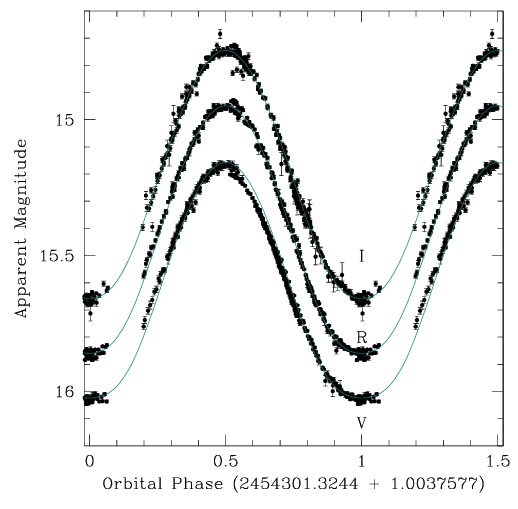}
    \includegraphics[width=7.2cm]{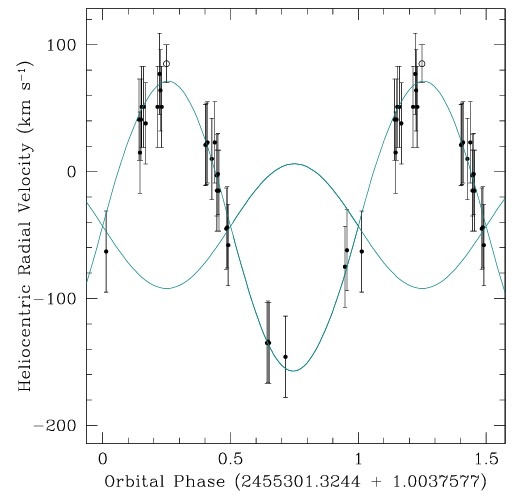}
    \caption{The light-curve (left panel; brightness vs. orbital phase) in three spectral bands (I, R and V) and radial velocity curve (right panel; stellar velocity along the line of sight caused by reflex motion to an orbiting companion) for the central star of the PN Abell~65 from figures 2 and 4 of \citet{Hillwig2015}. Modeling this information (blue curves) allows one to characterize the central binary with reasonable precision.}     
    \label{fig:lightcurve}
\end{figure}

Of the known binary central stars, several have been characterized. Most of these objects are very compact with periods less than 3 days and derive from a common envelope interaction \citep{Ivanova2013} between an AGB star and a companion that can be a main sequence star or a white dwarf. Their masses, and orbital parameters have been measured via modeling of the light curve and the radial velocity curve when available \citep[e.g.,][Figure~\ref{fig:lightcurve}]{Hillwig2004}. These studies have also showed that the orbital axis of the PN is aligned with the symmetry axis of the PN, a powerful conclusion to connect the morphology of the nebula to the binary interaction \citep{Hillwig2016}. 

For a complete review of multiplicity in PN see \citet{DeMarco2009b} and \citet{Jones2017}.

\section{Planetary Nebulae and their Parents, Siblings, Children, Cousins and Neighbors}
\label{sec:}

There are two reasons why we want to know about objects related to PNe. The first one is practical: some nebulae may contaminate the \textit{bona fide} PN sample with objects whose evolution is completely unrelated to that of PNe. The second reason is that some objects may not be exactly the same as PNe, but their existence, numbers and parameters may shed light on the evolution of PNe. By definition, PNe are nebulae surrounding the exposed nucleus of their progenitors, ionized by the nucleus' UV radiation. We can add that they derive from low-to-intermediate mass stars that {\it do not} go through a supernova explosion. Another way to define PNe is nebulae that derive from AGB stars, namely stars whose core is made of carbon and oxygen or oxygen and neon, having gone at least through helium burning. 

\subsection{Objects That Can Be Confused With PNe}
\label{ssec:objects_that_can_be_confused_with_pn}

The first type of ionized nebula that can be confused with PNe are those around symbiotic stars (a white dwarf ionizing the envelope/wind of its giant stellar companion), H~{\sc ii} regions (ionized gas in star-forming regions, where one or more massive stars with O or B spectral type ionizes interstellar gas), novae (white dwarfs that went through a surface thermonuclear detonation after accreting mass from a companion and ejected a nebula) or supernova remnants (ejecta of core-collapse or thermonuclear supernovae). 

From a purely observational perspective, several methods have been proposed to distinguish different ionized nebulae. \citet{GutierrezMoreno1995} presented an emission line diagnostic diagram for the separation of PNe from symbiotic nebulae based on the ratio of the emission lines [O~{\sc iii}] at 4363~\AA\ and H$\gamma$ versus [O~{\sc iii}] at 5007~\AA\ and H$\beta$, that reflects the difference  in electron densities between the two classes \citep{GutierrezMoreno1995,Clyne2015}. \cite{Ilkiewicz2017} presented a series of optical diagnostic diagrams to distinguish PNe from symbiotic nebulae, concluding that the [O~{\sc iii}], [N~{\sc iii}] and He~{\sc i} emission lines are more efficient due to their high critical densities and the notable different electron density between the two ionized nebulae. Other authors have also proposed diagnostic diagrams based on photometric data in the near- and mid-IR regimes \citep{Luud1987,Leedjarv1992,Schmeja2001,Phillips2007,Corradi2008,Akras2019b}.

\begin{figure}[h!]
    \centering
    \includegraphics[width=16.5cm]{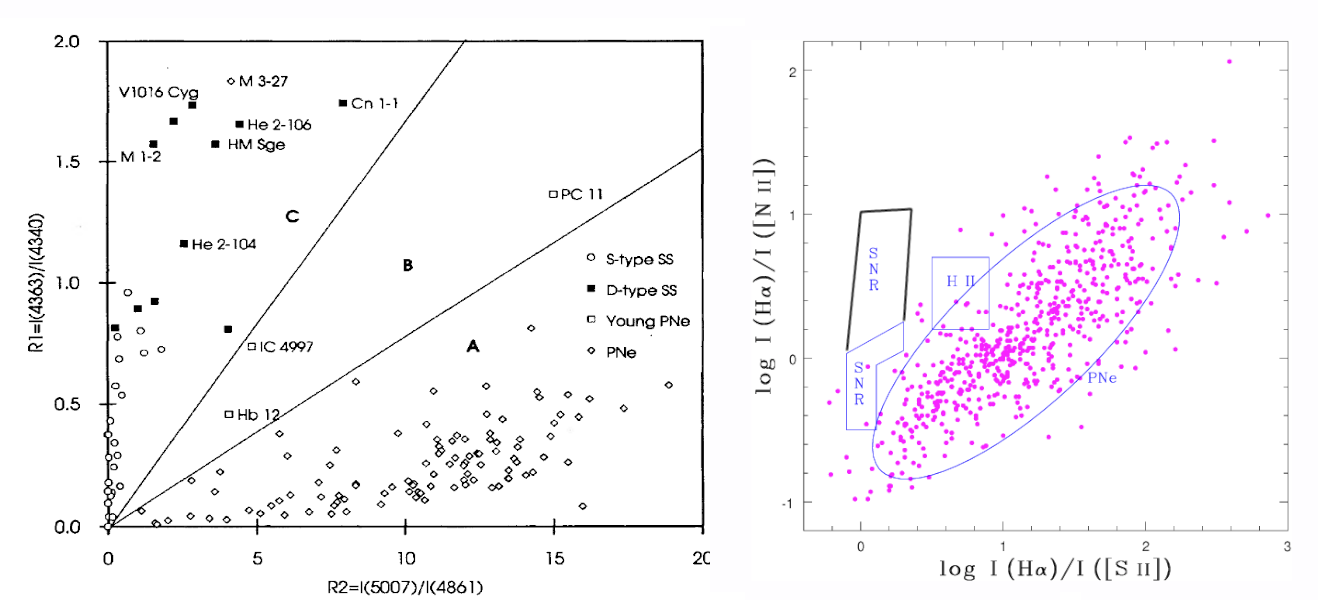}
    \caption{Diagnostic diagrams for distinguishing PNe from symbiotic stars, supernova remnants (SNR) and H~{\sc ii} regions. Left panel: The plot of line ratios [O~{\sc iii}] at 4363~\AA\  and H$\gamma$ versus [O~{\sc iii}] at 5007~\AA\ and H$\beta$ allows one to separate \textit{bona fide} PNe (region A) from symbiotic stars (S-type and D-type SS; region C). From figure 1 of  \citet{GutierrezMoreno1995}. Electron density increases from region A (low density gas), B (moderate density gas) to region C (high density gas). Right panel: The H$\alpha$/[N~{\sc ii}] versus H$\alpha$/[S~{\sc ii}] diagnostic diagram helps to separate PNe from SNRs and H~{\sc ii} regions. The ellipse indicates the probability density of 85\% that the object is a PN. Polygons indicate the regions where H~{\sc ii} regions and SNRs are found. The black area indicates the extended regime of SNRs for lower metallicity environments \citep{Leonidaki2013}. Figure modified from figure 5 of \citet{Riesgo2006}.}
    \label{fif:diagnostics}
\end{figure}

Supernova remnants can be confused with PNe, particularly when they are compact, as is the case for distant ones, such as those in other galaxies. They exhibit an emission line spectrum due to high-velocity shock waves resulting from their interaction with the interstellar medium, rather than from photo-ionization due to a hot central source. The distinction between PNe and SNRs is based instead on the well-known emission line diagnostic diagrams between the H$\alpha$/[S~{\sc ii}] and H$\alpha$/[N~{\sc ii}] ratio (right panel in Figure~\ref{fif:diagnostics}) \citep[e.g.,][]{Sabbadin1977,Riesgo2006,Frew2010,Leonidaki2013,Sabin2013,Akras2020}. The critical selection criteria for identifying shock-heated gas, and thus supernova remnants, is the [S~{\sc ii}]/H$\alpha>$0.4. Recently, \cite{Kopsacheili2020} demonstrated based on a machine learning approach that the [O~{\sc i}]/H$\alpha$ ratio provides a critical selection criterion for distinguishing supernova remnants from PNe.

The temperatures of the stars responsible for the heating and ionizing/exciting  the surrounding nebulae such as PNe and H~{\sc ii} regions is significantly different, the former have 30\,000~K$< T <$200\,000~K and the latter $T<$30\,000~K. This results in notable different ionization degrees between PNe and H~{\sc ii} regions, which is reflected in various line ratios. The majority of PNe are separated from H~{\sc ii} regions via the detection of the high ionization recombination line of He~{\sc ii} at 4686~\AA\ or the collisionally excited line [O~{\sc iii}] at 5007~\AA. In case of relative cold central stars, the low ionization degree PNe cannot be easily distinguished, especially the extragalactic ones.

\subsection{Objects That Are Related to PNe}

\begin{figure}
    \centering
    \includegraphics[width=7.0cm]{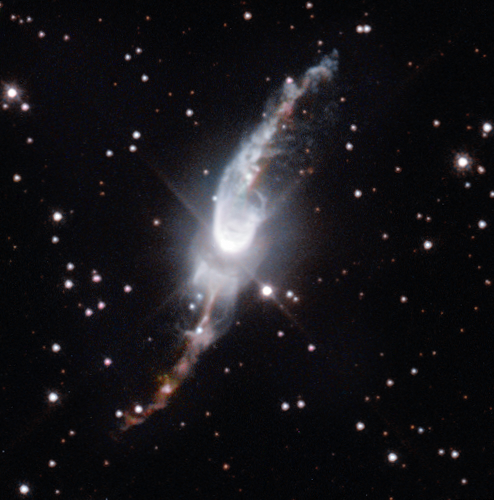}
    \caption{The pre-PN Hen~3-1475 imaged by the HST. This pre-PN is slightly bigger than a light year from one tip to the other. Credit: ESA, A. Riera (Universitat Politecnica de Catalunya, Spain) and P. Garcia-Lario (European Space Agency ISO Data Centre, Spain).}    
    \label{fig:PPN}
\end{figure}

There are many classes of objects related to PNe. To start there are  their immediate predecessor,  post-AGB stars, that have temperatures between 4000 and 10\,000~K. Some post-AGB stars have a pre-PN, a nebula that is not yet ionized because the central star has not yet warmed up to ionization temperatures (Figure~\ref{fig:PPN}). It is becoming apparent, that the pre-PN phenomenon may belong to a subset of post-AGB stars that have gone through a merger interaction with a companion. These nebulae are always collimated and when their outflow can be accurately measured it is clear that they have energies and momenta that imply a powerful engine at the source (and one that can also collimate the outflow). The only viable engine is the gravitational energy of the core, which can be mined effectively by a companion being disrupted and accreting onto the core \citep[][]{Sahai1998,Bujarrabal2001,Nordhaus2006,Blackman2014}. The fact that none of these stars was ever discovered to harbor a close companion argues for a merger. Some pre-PN central stars appear to have a wide companion with a period of years to decades \citep{Hrivnak2024}. In such cases the systems would have been triple stars.

Several post-AGB stars do not have a nebula at all. Of these, a fraction harbours intermediate-separation binary companions,  circumbinary disks and often jets that emanate from the companion \citep{2022A&A...666A..40B}.  It is not clear why the great majority of these post-AGB binary stars, with separation that indicate a recent binary interaction do not have a nebula \citep{vanWinckel2009}.

The descendants of PN central stars are white dwarfs - the PN having faded and recombined sufficiently to become unobservable. These stars take potentially very long times to cool and are visible (though faint and fading) for extremely long times (with the more massive ones cooling and fading faster than the less massive ones). There are scores for white dwarfs with different surface compositions due to their previous evolution as well as what elements float to the surface of their very low mass and thin atmospheres. Some white dwarfs are in binary systems, both compact as well as wide. The population properties of white dwarfs such as masses, composition, space distribution, etc. are related to the same properties in the PN population \citep{Tremblay2024}.

We have already labeled symbiotic star nebulae as look-alikes: unlike PNe, they are due to the wind of a giant companion ionized by a white dwarf. While these nebulae are not PNe, the  system that we call the symbiotic binary in the centre of the nebula is definitely related to binary central stars of PNe. Symbiotics are long period (few hundred to a few thousand days) interacting binary systems, comprising an evolved white dwarf and a less evolved giant that can be an RGB or an AGB star \citep[e.g.,][]{Belczynski2000,Mikolajewska2012,Akras2019a,Munari2019}. It is likely that these binaries were central stars of a PN when today's white dwarf had just left the AGB and today's giant companion, was a main sequence star, with a relatively long period. The progeny of symbiotic systems are double white dwarfs with long periods. 

Genuine PNe are considered as deriving from stars that have gone through evolution on the AGB. As such, their central stars have a carbon-oxygen core (or, more rarely an oxygen-neon core) and the PN gas is enriched with carbon as well as s-process elements such as barium, yttrium or strontium. With deeper PN surveys and larger systematic analyses, a handful of nebulae have been discovered that have central stars with masses that are clearly smaller than the helium-burning mass limit ($\sim$0.46~$M_\odot$; \citealt{Iben1995}). This means that these stars have not gone through helium burning and therefore cannot have evolved on the AGB (yet). These stars are instead post-RGB. They tend to be in binary systems \citep[e.g.][]{Jones2020}, where the mass of the central star is measured from modeling of the light and radial velocity curve of the compact (post-common envelope) binary. It is likely that the reason a nebula exists around these binaries, is that the two stars have interacted during the RGB evolution of the primary, and the relatively high and fast mass-loss event resulted in a visible nebula even if the post-RGB central stars with lower masses have smaller luminosity and are not particularly hot (though hot enough to ionized the circumstellar material).

\section{Planetary Nebulae as Tools to Study Cosmic Environments}
\label{sec:extragalactic_planetary_nebulae}

Planetary nebulae in specific environments, such as clusters or galaxies, can be used as tools to study the formation or evolutions of those environments. PNe embedded in populations of stars located all at approximately the same distance, also give us a tool to study PNe themselves, without the ever problematic distance determination. 

PNe are useful as chemical probes because of their easily measurable abundances. In particular, the abundance of certain elements that are not created or changed in PN central stars, give a clue as to the makeup of the original cloud from which the central star formed. PNe are also excellent kinematic probes in virtue of their sharp emission lines that allow us to measure the motion of the PN via the Doppler effect. 

PNe were classified into different types (I to IV) by \citet[][see also \citealt{Peimbert1980,FaundezAbans1987}]{Peimbert1978}. These types reflect specific abundances of elements like helium, carbon or oxygen, which in turn act as indicators of, e.g., the mass of the star. Different PN types can be useful for different measurements; for example Type II PNe are effective to determine the compositional gradients in galaxies, which in turn can give information about those galaxies' formation (see Section~\ref{ssec:tools_galaxies}). 

\subsection{Planetary Nebulae in Different Populations}\label{ssec:planetary_nebulae_in_different_populations}

PNe in globular clusters have been useful primarily as a tool to understand PNe rather than the clusters. This is because contrary to ``field" PNe, those that are scattered in our Galaxy, those in globular clusters have a known distance. The (approximately) single, old age of Galactic globular clusters means that all PN central stars should be of low mass (the higher mass ones transitioned though the PN phase earlier in the clusters' histories). The number of PNe in globular clusters also depends on the length of the transition time (the time it takes a PN to become ionized), as well as on the visibility time (the time it takes a PN to fade). These considerations were carried out by \citet{Jacoby2017}, who also realized that at least one of the four PNe in the Galactic globular cluster system was likely a merger of two stars, hence anomalous, one may not be a PN at all (IRAS~18333-2357) and one may not be in the original cluster at all \citep[JaFu~1;][]{Bond2020}. Despite a modern search using the MUSE instrument no additional PNe were found in the globular cluster system of the Galaxy \citep{Goettgens2019}, though some candidates were identified by \citet{Minniti2019}.

\begin{figure}
    \centering
    \includegraphics[width=10cm]{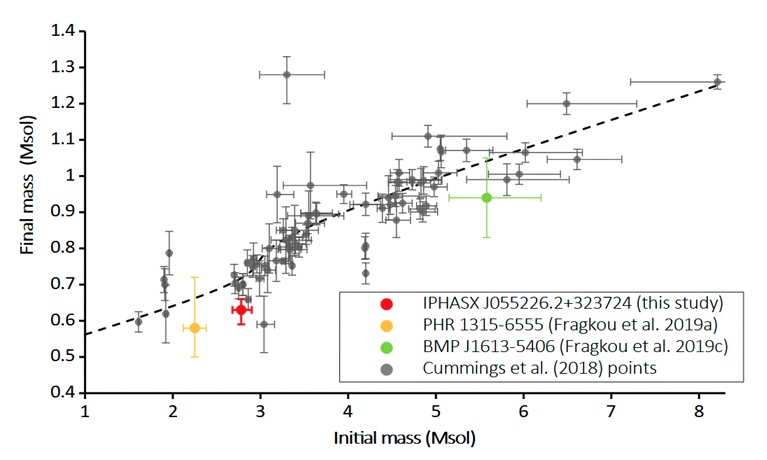}
    \caption{A plot from \citet[][from their figure 2]{Fragkou2022} of the known sample of cluster white dwarfs (grey circles) for the latest initial-to-final-mass estimates and semi-empirical model fit, together with estimated point for open cluster PNe IPHAS J055226.2+323724 (red circle), PHR~1315-6555 (yellow circle) and BMP~J1613-5406 (green circle). }    
    \label{fig:clusterWDs}
\end{figure}

There are three PNe in the open cluster system of the Galaxy, IPHASX~J055226.2+323724 in the open cluster M~37 \citep{Werner2023}, and two PNe in NGC~6067 and in Andrews-Lindsay 1 \citep[][Figure~\ref{fig:clusterWDs}]{Fragkou2022,Parker2011}. PNe in open clusters can help independently determine the initial-to-final mass relation for stars below 8~M$_\odot$, that connects the birth mass of a star to its white dwarf (final) mass, a critical measurement of how stars loose mass over their lives.

PNe have been used to study the Galactic bulge (the central region of the Galaxy). The formation of the Bulge remains mysterious with questions over when it formed and whether its formation was prolonged in time or whether it came abound relatively rapidly. The abundances of stars and nebulae in the Bulge can give clues as to when it formed and how long it took \citep{Barbuy2018}. In this context, PNe provide not just a sample of objects that have an approximately determinable age, but also have the ability to reveal with good accuracy the metallicity of the gas the star formed from. Their sharp lines also provide sensitive kinematic indicators showing the motion of various parts of the Bulge, like for example the ``bar" \citep{Beaulieu2000}.

\subsection{Planetary Nebulae as Tools to Understand Galaxies}\label{ssec:tools_galaxies}

PNe in external galaxies were first identified in the nearby Andromeda Galaxy \citep[M~31;][]{Baade1955}. Today PNe have been detected and measured in dozens of galaxies out to 40~Mpc (130 million light years; \citealt{Jacoby2024}). 

PNe in external galaxies can be used as kinematic tracers: their bright and sharp spectral emission lines allow us to measure their motion along the line of sight using the Doppler technique. This motion and their location within a given galaxy allow us to determine the overall motion of the PNe population in that galaxy. Using then the fact that most of the PN central stars have an intermediate age, we can assign an age to the kinematic structure that the PN population is part of. Typically, the motion of PNe in a galaxy is also compared to that of the interstellar gas, as well as that of other populations of stars with, for example, younger ages. Thus, one can build a map of galactic components, which have a known age and motion, and this helps clarify the formation history of galaxies \citep[as was done, e.g., for the galaxy NGC~6822;][]{FloresDuran2014}.

By measuring the abundances of PNe whose central stars have known mass adds more information regarding that galaxy's history. For example, more massive central stars tend to be younger than less massive ones. PNe around younger stars have gas that carries the chemical signature of the recent interstellar gas from which the PNe formed. There are many ways to interpret PN abundances in external galaxies, but the information that can be surmised has given rise to a rich and complex field of study \citep[for some modern references see, e.g.,][]{Bucciarelli2023}. PN abundances can also be compared with those of other populations of stars, which have different evolutionary histories and kinematic properties. Through these combined studies the history of the galaxy can be surmised, such as, for instance if and when a merger with another galaxy has taken place \citep[e.g.,][]{Bhattacharya2022}.

\subsection{Planetary Nebulae, the Hubble Constant and Cosmology}\label{ssec:Hubble_cte}

In the 70s and 80s the factor-of-two discrepancy between different measurements of the Hubble constant (between 50 and 100 km/s/Mpc) dominated many fields of Astronomy and Cosmology with big personalities dominating the science and little discourse that may favour intermediate values \citep{Jacoby1992}. Over the past two decades advances in observational techniques have reduced the uncertainty on measured values of the Hubble constant to a much smaller 1-2~\%. However, this increase in precision has come with its own problems as measurements of the Hubble constant carried out with different methods remain discrepant at the 3-5~$\sigma$ level between the Cepheid calibration of Type Ia supernovae \citep{Riess2022} and that from the microwave background \citep{Aghanim2020}. 
This new controversy, named the "Hubble tension" needs to be resolved or it might imply that the currently most accepted cosmological model \citep{Deruelle2018} has some missing physics.

\begin{figure}
    \centering
    \includegraphics[width=17cm]{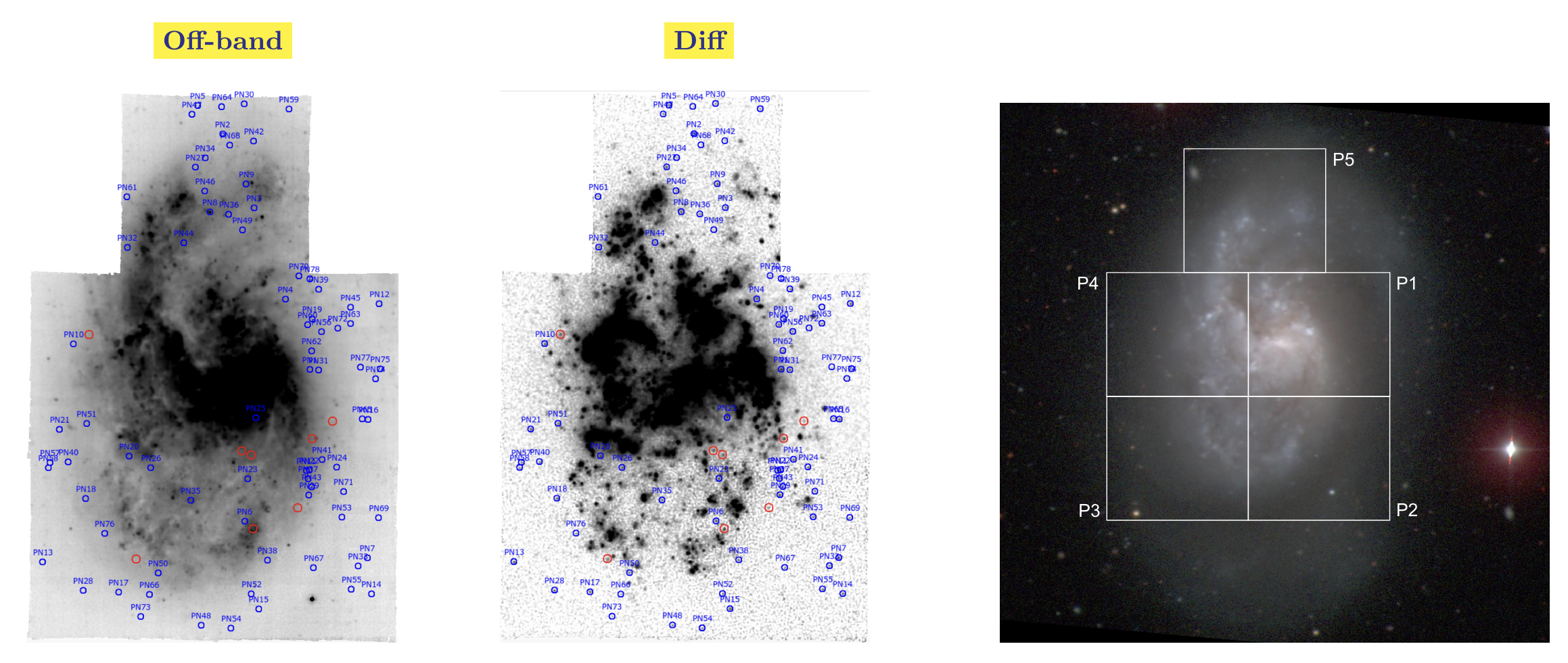}\\
    \includegraphics[width=7cm]{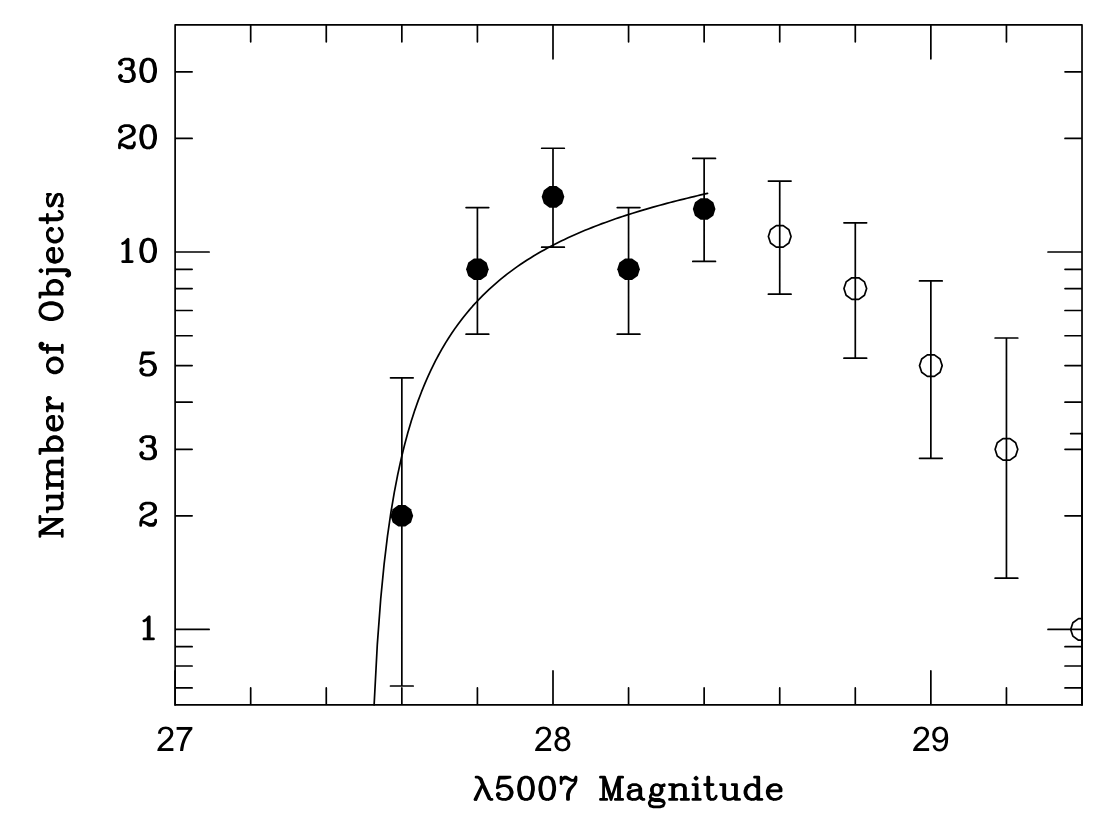}
    \caption{Top row: an off-band (left) and a difference image (middle) of the galaxy NGC~1385, highlighting the position of all the identified PNe (blue marks) and some contaminants (red marks). Right: a broadband image of the galaxy with an overlay of the MUSE fields. Bottom row: the PNLF for this galaxy binned in 0.2~mag bins. The error bars are 1$\sigma$ confidence interval. The black curve is a fit using Equation~\ref{eq:PNLF}. The open symbols are data beyond the completeness limit.  Figure from figures 18 and 19 of \citet{Jacoby2024}.}
    \label{fig:PNLF}
\end{figure}

Critical to any cosmological model is knowledge of distances to galaxies. This has been achieved with a series of interlaced methods that determine distances to objects with increasing distances - this set of methods are known as the ``distance ladder".  Within the context of the distance ladder, PNe have been used as standard candles to measure distances within the Local Group of galaxies. PNe can be easily identified out to large distances by their bright [O~{\sc iii}] lines, whose brightness can be measured accurately. Once a number of PNe are identified that belong to a given galaxy and are therefore equidistant from us, their [O~{\sc iii}] brightness can be histogrammed making a PN luminosity function (PNLF; Figure~\ref{fig:PNLF}). The PNLF can be fit with a function like:
\begin{equation}
    N(M) \propto e^{0.307M} (1 - e^{3(M-M_*)})
    \label{eq:PNLF}
\end{equation}
where $M$ is the [O~{\sc iii}] 5007~\AA\ magnitude while $M_* = -4.54$ is the brightest possible absolute magnitude in a PN. 

What is found is that the fitted bright end cut-off of the PNLF is invariant. Despite the lack of a theoretical underpinning for why PNe may have standard candle properties, and the notorious difficulty of measuring distances to Galactic objects, the techniques showed great promise \citep{Ciardullo1992}. 

The current explanation of why PNe exhibit a universal bright cut-off rests on the apparent fact that the more massive objects, which have brighter PNe, are also those that produce more dust, and where the dust has less time to disperse (in virtue of the short timescales over which these PN evolve). As such, they become intrinsically more extinct, which generates a feedback effect that makes more massive central stars with brighter PN invisible. Other explanations range from the universal presence of a group of blue straggler stars which are more massive because they are mergers \citep{Ciardullo2005}, to a group of stars where the luminosity of the bright end of the PNLF is powered by accretion onto a white dwarf \citep{Souropanis2023}. Independently of what mechanism generates bright PNe, it appears to be universal and to operate in old and young populations alike. As such the PNLF distance method appears to be robust.

A criticism to the PNLF distance method, stems from a number of possible contaminants - nebulae that can be mistaken for PNe but are not (see Section~\ref{ssec:objects_that_can_be_confused_with_pn}). Supernova remnants \citep{Kreckel2017}, H~{\sc ii} regions, background Ly$\alpha$ galaxies emitters \citep{Ciardullo2002b} can all be mistaken for PNe. These possible drawbacks can however all be addressed. Current measurements relying on the MUSE instrument on the Very Large Telescope, under strict quality control (small seeing, deep images, and ensuring that there are at least 50 PNe within 1 magnitude of the bright cutoff) have returned distances to individual galaxies with uncertainties that are a few percent of other known methods, showing great promise \citep{Jacoby2024} not only for the technique but for applying it in tandem with extremely large telescopes that will enable the PNLF to be used out to 2-3 times the current upper distance limit of $\sim$35~Mpc (about 100 million light years).

\section{Planetary Nebulae in the Era of Big Data}
\label{sec:planetary_nebulae_in_the_era_of_big_data}

When we talk about ``big data" in astronomy we think of surveys of millions of objects, stars or galaxies, that can then be studied with statistical techniques to reveal information that is not otherwise possible to surmise from smaller datasets with uncertain biases. In that sense PNe would not compete with other astronomical objects in that they are inherently a small population. This said, with the increasing number of surveys, automated techniques can discover more objects in a more systematic way, opening the possibility of more homogeneous targeted studies that are not biassed by brightness or PN size (large PNe tend to be old, faint and hard to detect). Below we list current efforts to create more complete samples. 

PNe were first listed in Messier's Catalogue of Nebulae and Star Clusters \citep{Messier_1781}, when the nature of these objects was still unknown. The first dedicated catalog of PNe was published by \citet{Perek_Kohoutek_1967_catalogue}. 
The Strasbourg-ESO Catalogue of Galactic Planetary Nebulae \citep{Acker_etal_1992_ESO_Catalogue} is another important PN catalogue, which lists 1143 object classified as true or probable PNe. In 2001, Kohoutek continued the original catalog, increasing the number of known PNe to 1510 \citep{Kohoutek_2001_Catalogue}.

Nowadays, the count of known Galactic PNe has rapidly increased to nearly 3800, as documented in the Hong Kong/AAO/Strasbourg/H$\alpha$ PN catalogue \citep[HASH;][]{Parker2006,Parker2016}. Nevertheless, this figure is likely lower than the expected from, for example, counting PN in external galaxies. The latter figure is of the order of 10\,000 members, but it depends on the observability limit of PN surveys. For a discussion of these estimates see \citet{Moe2006}.

The noteworthy surge in PN numbers over the past 15 years can be attributed to the advent of large-scale photometric surveys covering large portions or even the entirety of the  sky. These surveys provide information for hundreds of millions of objects of different types. Examples are the SuperCOSMOS AAO/UKST H$\alpha$ Survey (SHS) of the Southern Galactic plane \citep{Parker2005} and the Isaac Newton Telescope Photometric H$\alpha$ Survey (IPHAS) of the Northern Galactic plane \citep{Drew2005}. Additionally, new genuine PN discoveries have also emerged from other, non-dedicated surveys, such as the UWISH2 survey 
\citep[183;][]{Gledhill2018}, the CORNISH survey (90; \citealt[][]{Irabor2018} and 62; \citealt[][]{Fragkou2018}) or a number of {\it ad hoc} surveys \citep[e.g., 48;][]{Boumis2003,Boumis2006}.

However, traditional methods for discovering new PNe are struggling to keep pace with the amount of data provided by new surveys, necessitating the development of more sophisticated approaches. In recent years, advanced automated machine or deep learning algorithms and artificial neural networks have been employed to explore publicly available data. A decision tree algorithm achieved a 75~\% accuracy in distinguishing compact PNe from other H$\alpha$ emitters such as H~{\sc ii} regions, by combining optical and IR photometric data \citep{Akras2019c}. Subsequently, \cite{AwangIskandar2020} utilized three deep learning algorithms to automatically separate PNe from other sources, combining images from the HASH and Panoramic Survey Telescope and Rapid Response System (Pan-STARRS; \citealt{Kaiser2002}) databases. The accuracy of their models ranged between 60 and 90~\%, depending on the algorithm and image training sample. 

Finally, \cite{Sun2024} employed advanced artificial intelligence and deep learning algorithms to automatically detect PNe in large image databases, such as the VPHAS+ catalog \citep{Drew2014}. This endeavor yielded a list of 815 candidate PNe, and subsequent spectroscopic follow-up observations of 31 candidates unveiled the discovery of 23 new PN members.

Despite these advancements, many more PNe remain undiscovered, necessitating further exploration of databases from completed and ongoing surveys, such as the Southern Photometric Local Universe Survey (S-PLUS; \citealt{MendesdeOliveira2019}), Javalambre Photometric Local Universe Survey (J-PLUS; \citealt{Cenarro2019}), the Javalambre Physics of the Accelerating Universe Astrophysical Survey (J-PAS; \citealt{Benitez2014}), the Large Synoptic Survey Telescope (LSST; \citealt{Ivezic2019}), the Large Sky Area Multi-Object Fiber Spectroscopic Telescope (LAMOST; \citealt{Cui2012}) spectroscopic survey, or the GAIA photometric and spectroscopic data \citep{Prusti2016}.

\section{Planetary Nebulae in the Era of Big Telescopes}
\label{sec:planetary_nebulae_in_the_era_of_big_telescopes}

What can ultra high spatial resolution observations (in the milli-arcsec regime) of large samples of PNe in the mid-IR wavelength, ultra sensitive radio observations or even the detection of particle bursts bring to our understanding of PN? The new and upcoming observational platforms that can advance our knowledge of PNe in the next decade are likely to be the James Webb Space Telescope (JWST), the Square Kilometer Array (SKA) and its precursors such as the SKA Pathfinder (ASKAP), the ALMA, the Extremely Large Telescopes (ELT) and even the Cherenkov Telescope Array (CTA). Finally transient surveys (including the Large Survey of Space and Time, LSST, with the Rubin Observatory) will be able to catch outbursts in the middle of PNe, or those outbursts that will generate a PN in the future.

It has long been known that PN environments are extraordinarily rich chemical factories, where elements created in the stellar interiors are ejected at different times and in different directions with a multitude of speeds. The harsh environments produced by the quickly heating central star, lead to the formation of compounds that are constantly destroyed and reformed. Organic species are manufactured as well as dust grains that eventually make it into new generations of stars and planets. This chemical factory has been studied for decades, particularly since the launch of HST and since the first IR (ISO) and UV telescope (IUE and again HST) observations. However, the quality and quantity of the observations is not sufficient to constrain complex chemical scenarios. Fortunately, all this is already changing with JWST and ALMA, primarily, and will see new information in the era of the ELT.

Transient surveys will detect phenomena like novae and even supernovae that happen in the middle of PNe and even outbursts that create new PNe (which however will not become detectable for years to decades), like luminous red novae. Some of these may even generate cosmic rays detectable with new platforms like the CTA - although arguably this would be a much rarer event.

\begin{figure}
    \centering
    \includegraphics[width=15cm]{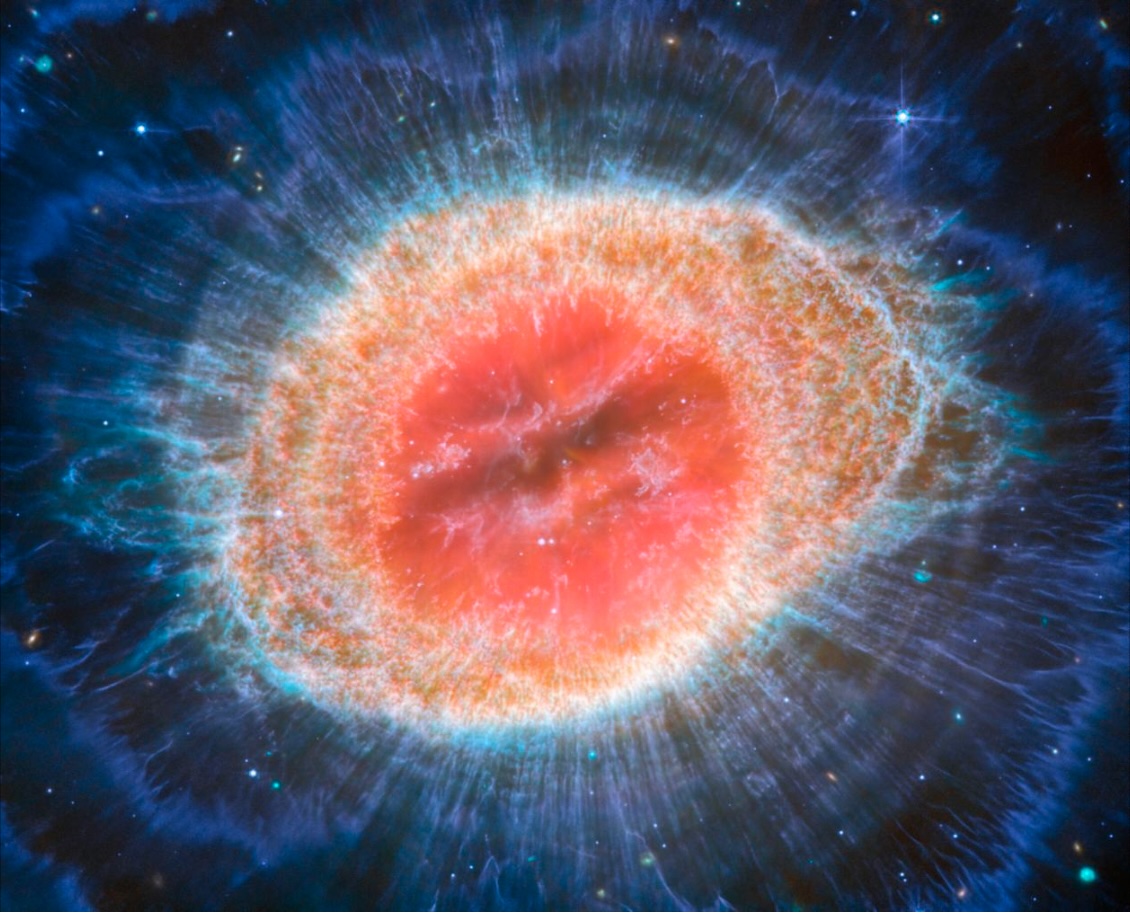}\\
    \caption{The Ring PN (NGC~6720) as observed by JWST's instrument MIRI using various filters. The image is approximately 1.5 light years across for a distance to the PN of 2300 light years. Credits: ESA/Webb, NASA, CSA, M. Barlow, N. Cox, R. Wesson.} 
    \label{fig:Ring}
\end{figure}

Examples of the exquisite level of detail easily achieved by JWST observations are the images of the Southern Ring Nebula (NGC~3132; \citealt{DeMarco_etal_2022}) and the Ring Nebula (NGC~6720; \citealt{Wesson2024}, see Figure~\ref{fig:Ring}). In these images the location and amount of molecular hydrogen (H$_2$) can be compared with the location of dust, giving early clues as to the relationship between the formation of H$_2$ and that of dust grains. Another example shows how distant and faint PNe can be now observed with similar levels of precision to Galactic ones. \citet{Jones_etal_2023} showed JWST observations of a PN in the Large Magellanic Clouds with 41 additional emission line measurements compared to the previously observed 10 lines. This promises to extend the use of PNe as extra-galactic and distance scale probes many folds.

\section{Conclusion}
\label{sec:conclusion}

Planetary nebulae are a crucial stage in the life cycle of stars with intermediate mass. They also play a crucial role in the the chemical evolution of galaxies by enriching the interstellar medium with heavy elements, forged in the stellar interior, that are essential for the creation of new stars and planetary systems. In this chapter, we have examined the processes that lead to their formation. We also explored the connection between PNe and binary systems, which are likely critical for the formation of their complex shapes. Additionally, we reviewed the significant advancements in modeling the interiors of PN's central stars over the past 50 years. We also discussed studies on the hot and cool gas and dust typically seen in PN across the electromagnetic spectrum. Finally, we highlighted the advantages of observing and studying planetary nebulae in other galaxies to enhance our understanding of galactic evolution and improve distance estimates on cosmological scales, as well as the exciting new opportunities presented by large facilities and advanced telescopes.

\begin{ack}[Acknowledgments]

  SA acknowledges support from the H.F.R.I. call \lq\lq~Basic research financing (Horizontal support of all Sciences)\rq\rq~under the National Recovery and Resilience Plan \lq\lq Greece 2.0\rq\rq~funded by the European Union – NextGenerationEU (H.F.R.I. Project Number: 15665).

  We have made use of NASA's Astrophysics Data System Bibliographic Service.

\end{ack}

\bibliographystyle{Harvard}
\bibliography{reference}

\end{document}